\documentclass[conference]{IEEEtran}
\IEEEoverridecommandlockouts
\usepackage{cite}
\usepackage{amsmath,amssymb,amsfonts}
\usepackage{algorithmic}
\usepackage{graphicx}
\usepackage{textcomp}
\usepackage{xcolor}
\usepackage{algorithm}
\usepackage{diagbox}
\usepackage{multirow}
\usepackage{float}
\usepackage{bigstrut}
\usepackage{booktabs}
\usepackage{wrapfig}

\usepackage[switch]{lineno} 
\usepackage{amsthm}

\def\BibTeX{{\rm B\kern-.05em{\sc i\kern-.025em b}\kern-.08em
    T\kern-.1667em\lower.7ex\hbox{E}\kern-.125emX}}
\begin{document}
\title{SGDP: A Stream-Graph Neural Network Based Data Prefetcher}

\author{

\IEEEauthorblockN{Yiyuan Yang\footnotemark*}
\IEEEauthorblockA{\textit{University of Oxford} \\
\textit{Huawei Noah's Ark Lab}\\
Oxford, UK \\
yiyuan.yang@cs.ox.ac.uk}

\\

\IEEEauthorblockN{Gang Hu}
\IEEEauthorblockA{\textit{Huawei} \\
Chengdu, China \\
hugang27@huawei.com}

\and

\IEEEauthorblockN{Rongshang Li\footnotemark*}
\IEEEauthorblockA{\textit{University of Sydney} \\
\textit{Huawei Noah's Ark Lab}\\
Sydney, Australia \\
roli5128@uni.sydney.edu.au}

\\

\IEEEauthorblockN{Xing Li}
\IEEEauthorblockA{\textit{Huawei Noah's Ark Lab} \\
Hongkong, China \\
li.xing2@huawei.com}

\and

\IEEEauthorblockN{Qiquan Shi}
\IEEEauthorblockA{\textit{Huawei Noah's Ark Lab}\\
\textit{Huawei}\\
Shenzhen, China \\
shiqiquan@huawei.com}

\\

\IEEEauthorblockN{Mingxuan Yuan}
\IEEEauthorblockA{\textit{Huawei Noah's Ark Lab}\\
Hongkong, China \\
Yuan.Mingxuan@huawei.com}

\and
\IEEEauthorblockN{Xijun Li}
\IEEEauthorblockA{\textit{MIRA Lab, USTC} \\
\textit{Huawei Noah's Ark Lab}\\
Shenzhen, China \\
xijun.li@huawei.com}

}

\maketitle
\renewcommand{\thefootnote}{\fnsymbol{footnote}} 
\footnotetext[1]{Both authors contributed equally to this research. Work done as interns in Huawei Noah's Ark Lab.} 

\begin{abstract}
Data prefetching is important for storage system optimization and access performance improvement. Traditional prefetchers work well for mining access patterns of sequential logical block address (LBA) but cannot handle complex non-sequential patterns that commonly exist in real-world applications. The state-of-the-art (SOTA) learning-based prefetchers cover more LBA accesses. However, they do not adequately consider the spatial interdependencies between LBA deltas, which leads to limited performance and robustness. This paper proposes a novel Stream-Graph neural network-based Data Prefetcher (\textbf{SGDP}). Specifically, SGDP models LBA delta streams using a weighted directed graph structure to represent interactive relations among LBA deltas and further extracts hybrid features by graph neural networks for data prefetching. We conduct extensive experiments on eight real-world datasets. Empirical results verify that SGDP outperforms the SOTA methods in terms of the hit ratio by 6.21\%, the effective prefetching ratio by 7.00\%, and speeds up inference time by 3.13$\times$ on average. Besides, we generalize SGDP to different variants by different stream constructions, further expanding its application scenarios and demonstrating its robustness. SGDP offers a novel data prefetching solution and has been verified in commercial hybrid storage systems in the experimental phase. Our codes and appendix are available at https://github.com/yyysjz1997/SGDP/.

\end{abstract}


\section{Introduction}

In the big data era, the demand for high-performance storage systems is increasing rapidly. The Input/Output (I/O) speed gap between different storage devices in a hybrid storage system might cause high access latency \cite{kim2010flashfire}. To fill this gap, the cache is designed to temporarily keep data that are likely to be accessed in the future. The performance of cache, commonly represented by \textbf{hit ratio}, has a direct impact on the performance of the whole storage system.

To improve the hit ratio, data prefetching is introduced as an essential technique in the cache. Prefetchers reduce access latency by fetching data from their original storage in slower memory to cache before they are needed \cite{cachememory}. Common block-level cache prefetchers take in logical block address (LBA) access sequences as input (i.e., some integer numbers). Prefetchers predict the LBA of the block that might be accessed in a short time and decide whether to pre-load it or not. There are two major challenges in the design of effective prefetchers. First, the LBA access sequences in real-world applications have complex patterns due to concurrent and random accesses from different users or applications, which are common in modern large-scale storage systems \cite{basak2016storage,LI2021CFTT}. Second, effective prefetchers need to be accurate. Inaccurate prefetchers waste both I/O bandwidth and cache space \cite{hashemi2018learning}. Therefore, designing effective prefetchers is vital for storage systems.

Traditional prefetchers prefetch the data by matching LBA access sequences to specific predefined rules. However, they can hardly adapt to complex real-world scenarios as their predefined rules are limited to specific simple patterns such as sequential reading  \cite{boboila2011performance}. To learn complex patterns, several learning-based methods   \cite{liao2009machine,laga2016lynx,chen2021revisiting, wu2021survey} are applied. Recently, long short-term memory (LSTM) based methods like DeepPrefetcher \cite{ganfure2020deepprefetcher} and Delta-LSTM \cite{chakraborttii2020learning} have shown promising results. They model the \textbf{LBA delta} (i.e., the difference between successive access requests), which covers more LBA accesses. However, due to concurrent accesses, the chronological order of LBA deltas within a short time period is likely to be disrupted. DeepPrefetcher and Delta-LSTM disregard the internal temporal correlation and result in limited performance.

\begin{figure*}[tbp!]
	\centering
    \includegraphics[width=0.95\textwidth]{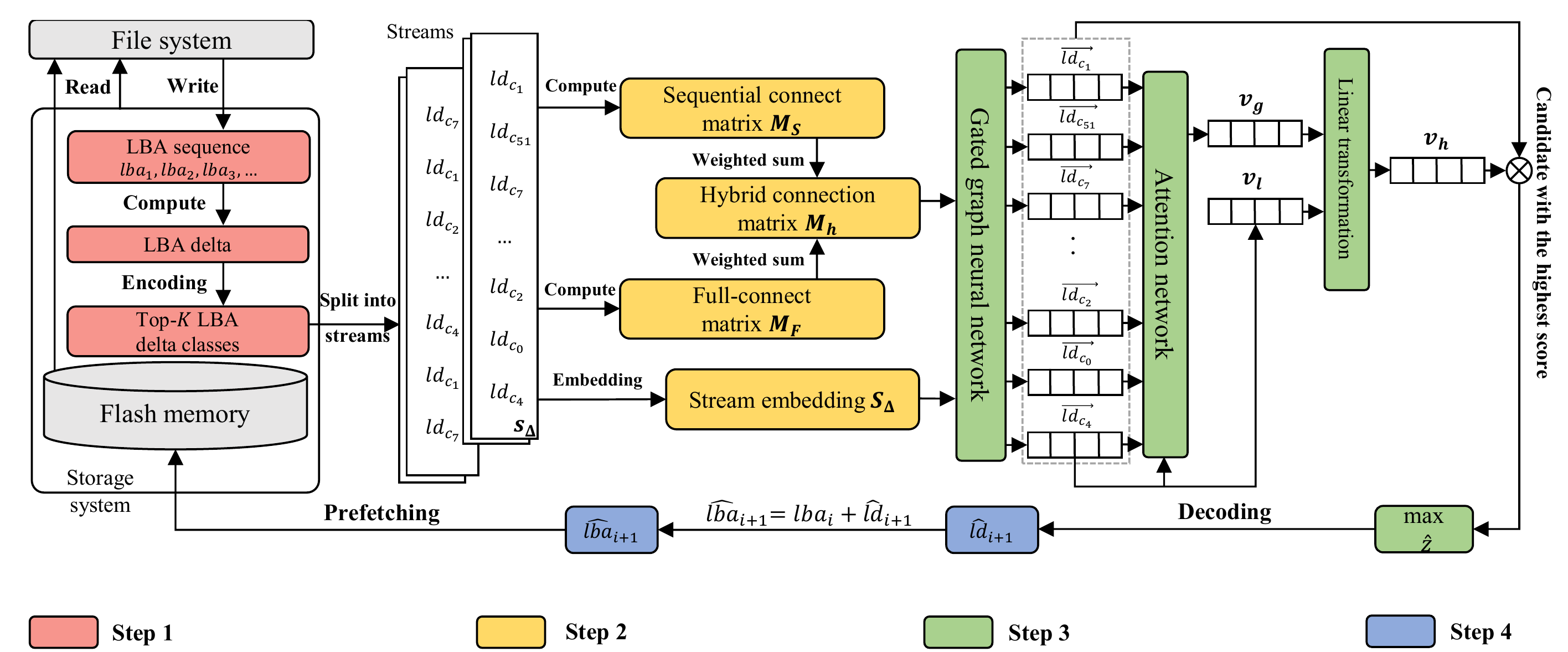}
    \caption{The workflow of the SGDP framework. In \textcolor[RGB]{232,153,143}{\textbf{Step 1}}, we compress the search space and reduce the learning complexity. In \textcolor[RGB]{248,217,120}{\textbf{Step 2}}, we compute the hybrid connection matrix $\textbf{M}_h$ with sequential and global information and embed the LBA delta stream into a matrix $\textbf{S}_\Delta$. In \textcolor[RGB]{176,207,148}{\textbf{Step 3}}, using gated graph neural networks to update the latent node vectors. Each stream is represented as the combination of the local preference $\textbf{v}_l$ and global interaction $\textbf{v}_g$ by an attention network. In \textcolor[RGB]{147,168,215}{\textbf{Step 4}}, we predict the candidate with the highest score and decode it to get the next accessed LBA for prefetching. This framework corresponds to the four steps of \textbf{Algorithm \ref{SGDPalg}}.}
    \label{fig_0}
\end{figure*}

Graph structures can effectively use nodes and edges to represent LBA (delta) and access sequence, and can mine intrinsic access patterns beyond chronological order in hybrid storage systems like relational databases. Therefore, to improve the performance of prefetching especially in applications with complex patterns, this work models the relations among LBA deltas using graph neural network, and proposes a novel method called \underline{S}tream-\underline{G}raph Neural Network-Based \underline{D}ata \underline{P}refetcher (\textbf{SGDP}), as shown in \textbf{Figure} \ref{fig_0}. Specifically, we encode LBA deltas and split them into shorter streams. Then we build weighted directed graphs based on LBA delta streams and extract relations of sequential connection and temporal accesses of LBA deltas from each stream, which are represented as sequential connect matrices and full-connect matrices, respectively. By fusing those two matrices, we get hybrid matrices that contain the relations of LBA deltas. Finally, the hybrid matrix, along with embedding LBA deltas of each stream is fed into a gated graph neural network to learn access patterns for prefetching. Extensive experiments on eight real-world datasets show that SGDP outperforms the SOTA prefetchers in terms of performance and efficiency.

The contributions of this work are summarized as follows:

1. SGDP can accurately learn complex access patterns by capturing the relations of LBA deltas in each stream. The relations are represented by sequential connect matrices and full-connect matrices using graph structures.

2. To the best of our knowledge, SGDP is the first work that utilizes the stream-graph structure of the LBA delta in the data prefetching problem. Using gated graph neural networks and attention mechanisms, we extract and aggregate sequential and global information for better prefetching.

3. As a novel solution in the hybrid storage system, SGDP can be generalized to multiple variants by different stream construction methods, which further enhances its robustness and expands its application to various real-world scenarios. 

4. SGDP outperforms SOTA prefetchers by 6.21\% on hit ratio, 7.00\% on effective prefetching ratio, and speeds up inference time by 3.13$\times$ on average. It has been verified in commercial hybrid storage systems in the experimental phase and will be deployed in the future product series.

\section{Related Work} \label{graphbased}

\textbf{Traditional Data Prefetcher} The most commonly used prefetcher is the Stride prefetcher \cite{fu1992stride} which uses a reference prediction table to store the last few accessed LBAs and the stride to obtain the required LBA. Although it can capture a constant stride in sequential access patterns, it can hardly detect variable strides in irregular access patterns. Temporal prefetchers learn irregular access patterns by memorizing pairs of correlated LBAs \cite{li2004c,wenisch2008temporal,wu2019temporal,wu2019efficient}. However, due to inconsistent correlation address pairs, these traditional methods cannot achieve good performance in practice.

\textbf{Learning-based Data Prefetcher} Prefetching needs to be accurate, as a small error in the numerical value of a prefetched LBA leads to useless prefetch and a waste of cache space and I/O bandwidth. Even though prefetching can be treated as a prediction problem, regression-based time-series prediction models like ARIMA are not widely considered by researchers. Classification-based methods seem more favoured because they can cover more LBA accesses, but not practical as it is hard to cover all LBAs. For example, in Microsoft Research Cambridge traces \cite{lee2017understanding}, the top-1000 most frequently occurring \textbf{LBAs} cover only 2.8\% of all the LBA accesses, whereas the top-1000 most frequently occurring \textbf{LBA deltas} cover 91.7\% of all LBA accesses \cite{chakraborttii2020learning}. To learn more complex access patterns, many deep learning approaches are proposed and consider the LBA delta as input directly  \cite{hashemi2018learning}. DeepPrefetcher \cite{ganfure2020deepprefetcher} transforms the LBA sequence into LBA deltas, then employs the word2vec model and LSTM architecture to capture the hidden feature in the input sequence. However, it is inefficient on large-scale trace datasets. Delta-LSTM is proposed \cite{chakraborttii2020learning} and addresses the large and sparse LBA space by co-learning top-$K$ LBA delta and I/O size features. Within top-$K$ (e.g., $K=1000$) classes, the searching space is restricted, which alleviates the class explosion problem. However, both of them do not consider the relations of LBA deltas to capture the more complex patterns (e.g., the continuous LBA accesses across many pages), which results in limited performance.
 
\textbf{Prefetching with Graph-based Structure} %
Complex patterns in LBA access streams can be constructed by graphs \cite{liao2015prefetching, zhu2020ctdgm}. Nexus uses metadata relationship graphs to assist prefetching decision-making  \cite{2006Nexus}. Ainsworth et al. design a prefetcher for breadth-first searches on graphs \cite{graph2016}. These methods transform a sequence of observed LBAs into a directed graph, in which a node is utilized to represent a block access event to model block access patterns. However, LBA accesses are quite sparse, which results in large graphs in these LBA sequence-based methods and makes these prefetchers quite ineffective in practice.

\section{Preliminaries}
Consider an LBA access sequence with length $n$:
\begin{equation}
\langle lba_i \rangle_{i=1}^{n} = \langle lba_1,lba_2,\ldots,lba_n \rangle,
\end{equation} 

\noindent in which $lba_i\in N$ represents the address number of the $i$-th accessed blocks. The data prefetching problem can be regarded as given $\langle lba_i \rangle_{i=1}^{n}$, predict $lba_{n+1}$. Following the previous learning-based prefetchers, we compute LBA deltas ($ld$):
\begin{equation}
   ld_i = lba_{i+1}-lba_{i},
\end{equation}
\begin{equation}
    \langle ld_i \rangle_{i=1}^{n-1} = \langle ld_1,ld_2,\ldots,ld_{n-1} \rangle.
\end{equation}

In order to get the prediction of the next LBA, $\widehat{lba}_{n+1}$, we predict the delta $\widehat{ld}_{n}$. Note that the variables with the $\hspace{0.1cm} \widehat{}\hspace{0.1cm}$ symbol denotes the predicted values. In short, the data prefetching problem is formulated as follows:
\begin{equation}
    \widehat{lba}_{n+1} = lba_{n} + \widehat{ld}_{n}.
	\label{Eq_0} 
\end{equation}

To restrict the model size, the number of classes of LBA deltas that needs to be predicted is capped to a fixed number of $K+1$. Here $K$ is acquired by top-$K$ most frequently occurring LBA delta in $\langle ld_i \rangle_{i=1}^{l-1}$, and the extra class is for the other infrequent LBA deltas. The prefetcher treats this extra class as no-prefetch because infrequent LBA delta is hard to predict. We redefine these $K+1$ classes as $LD = \{ld_{c_0}, ld_{c_1},\ldots,ld_{c_{K+1}}\}$, $ld_{c_0}$ representing the no-prefetch class and $\{ ld_{c_1},\ldots,ld_{c_{K+1}}\}$ representing the top-$K$ ones. The model predicts the $ld_{c_i}$ of the next LBA delta, and prefetches it when the predicted class is in the top-$K$ and does not prefetch if the class is $ld_{c_0}$. Using the LBA delta and Top-$K$ mechanism can effectively reduce the sparse problem and the search space of the model.

\section{Methodology}\label{SGDP_framework}

\subsection{LBA Delta Streams}\label{lbadelta}
It would yield expensive costs if building the directed graph of the LBA delta sequence generated by the whole LBA sequence (length $\ge 1\times10^7$). To alleviate this problem, we use the concept of data access \textbf{stream}. We split the whole LBA delta sequence into shorter streams which represent the temporal access patterns. Specifically, we consider LBA accesses with close access times to be in the same stream. Whenever the time interval between two LBA accesses is longer than a preset time limit $T$ (e.g. $T =$ 0.1ms), the LBA sequence will be split to generate a new stream. We then split the LBA delta sequence correspondingly and use a sliding window inside each stream to generate equal-length LBA delta streams. A split LBA delta stream $\textbf{s}_{\Delta}$ can be represented by a vector $\textbf{s}_{\Delta}= [ld_{s,1},ld_{s,2}, ..., ld_{s,n}]$ in the chronological order, where $ld_{s,i}$ denotes the $i$-th LBA delta in the stream $\textbf{s}_{\Delta}$. The LBA delta stream is not only suitable for building directed graphs but also implies the temporal locality of LBA accesses.

\begin{figure}[!t]
	\centering
    \includegraphics[width=0.495\textwidth]{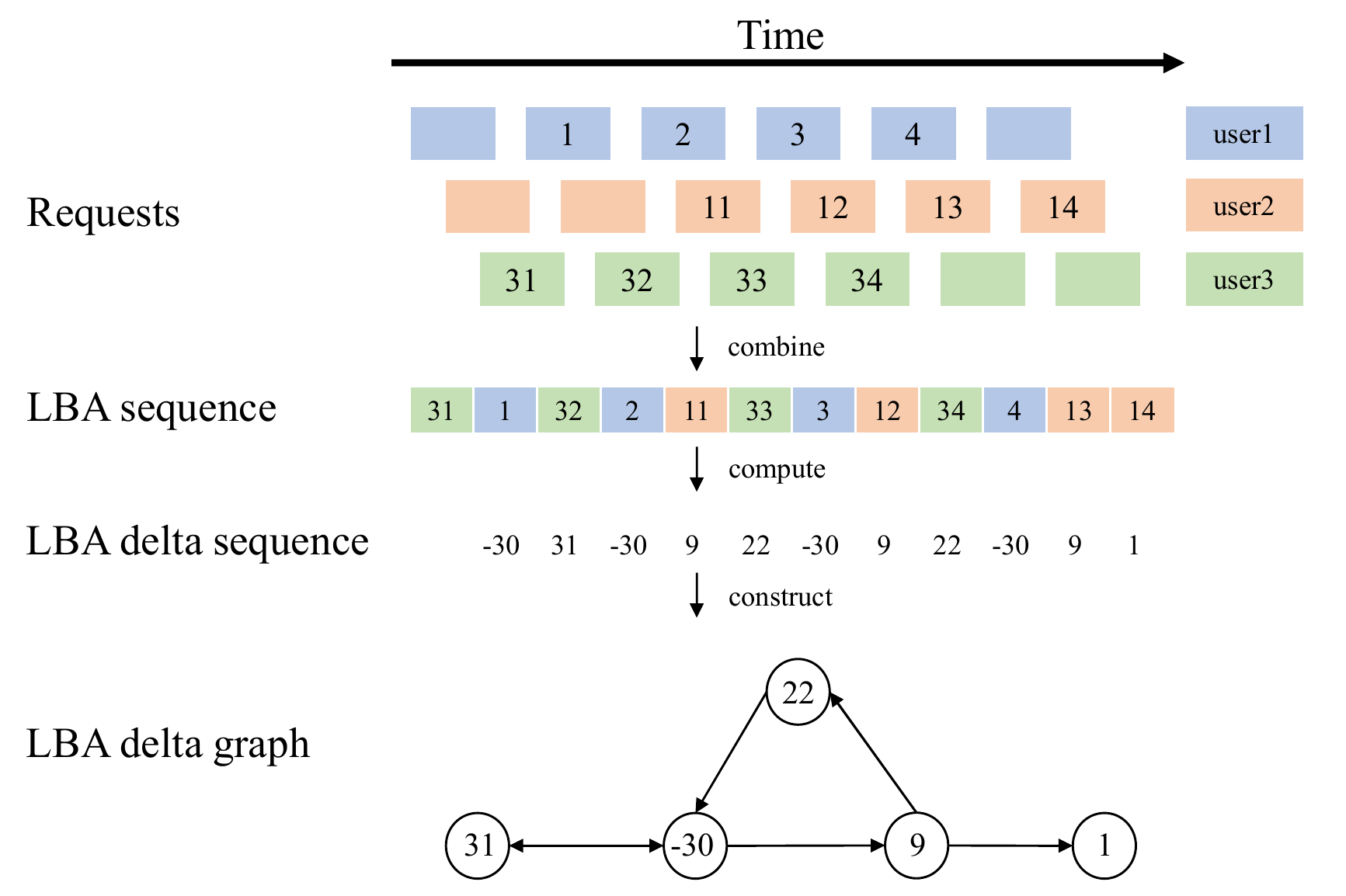}
    \caption{Example of LBA delta and graph.}
    \label{fig_2}
\end{figure}

\subsection{LBA Delta Based Graph Structure}




As discussed in \textbf{Section} \ref{graphbased}, LBA-based graph structure is ineffective for data prefetching in practice. To solve this problem, we propose a high-order graph based on the LBA delta. Here we present a toy example in \textbf{Figure} \ref{fig_2} to show that LBA deltas can also be represented by a directed graph. Consider three users sending concurrent sequential read requests. The concurrent requests are combined into one LBA sequence before being sent to the storage system. Following the LBA delta computation progress in \textbf{Section} \ref{lbadelta}, we can simplify the original LBA sequence that has 12 different LBAs and represent it with an LBA delta sequence with only 5 different nodes. We can build a directed graph based on the LBA delta sequence by using LBA deltas as a node and linking all the LBA deltas with their successor.

In contrast to previous works on extracting useful features from access patterns between nearby accessed LBAs only, we observed if an LBA request sequence is divided into several streams, different streams are possible to be accessed by a similar pattern. Non-sequential features can be extracted from the observed access patterns by monitoring the change (or difference) between successive LBA requests. This is achieved by learning the hybrid connection matrix (which contains adjacent and interactive relations) along with the embedded LBA deltas. Also, the number of the graph nodes is constant, that is $K+1$.

\subsection{Weighted Directed Stream-based Graph}

\begin{figure}[t]
	\centering    \includegraphics[width=0.472\textwidth]{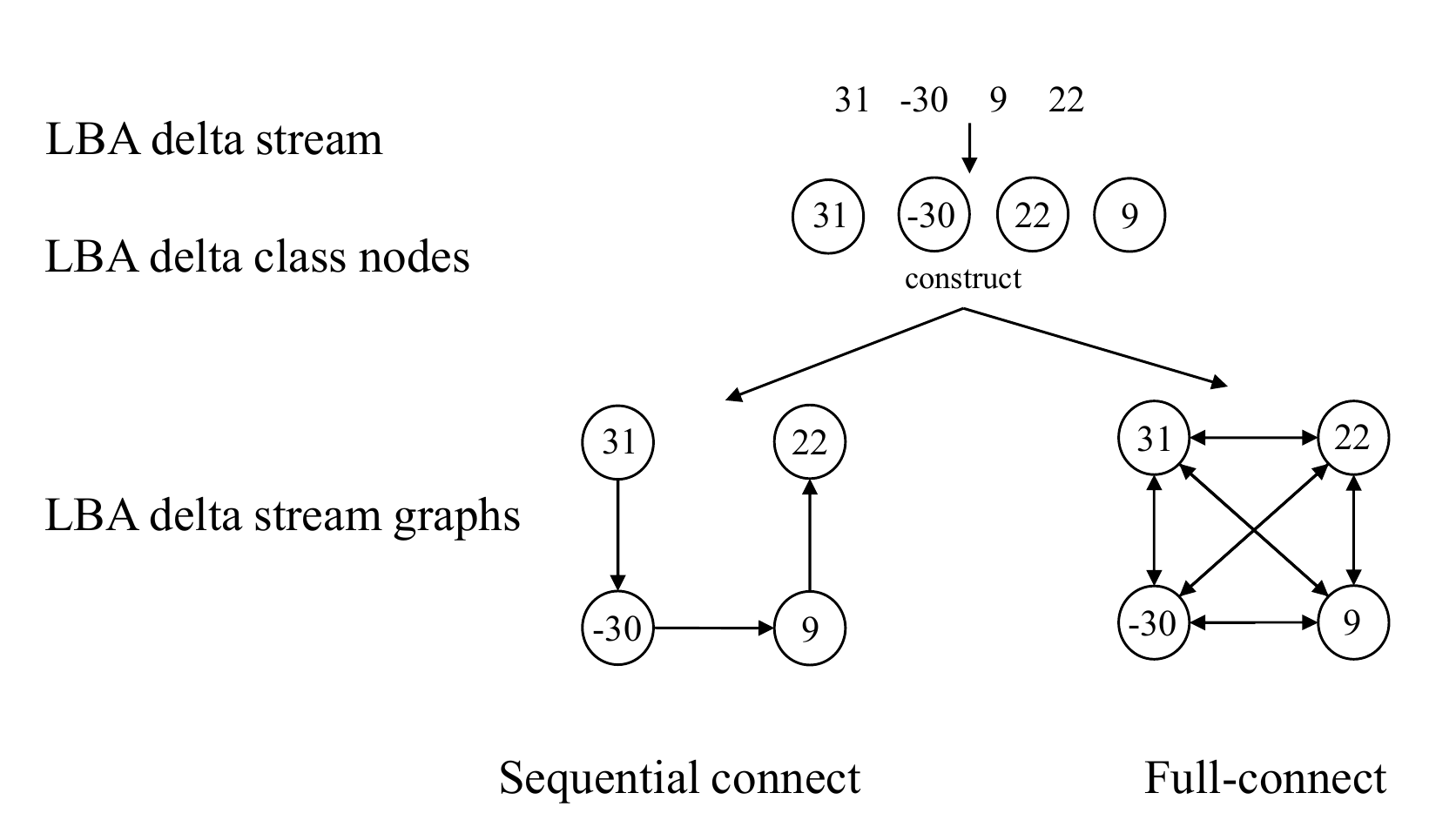}
    \caption{Example of two kinds of graphs.}
    \label{fig_3}
\end{figure}

Without loss of generality, each LBA delta stream $\textbf{s}_{\Delta}$ can be modeled as a directed graph $\mathcal{G}_{\textbf{s}_{\Delta}} = (\mathcal{V}_{\textbf{s}_{\Delta}},\mathcal{E}_{\textbf{s}_{\Delta}})$. Each node in the directed graph $\mathcal{G}_{\textbf{s}_{\Delta}}$ expresses one of the classes of LBA deltas $ld_{s,a} \in LD$. We build the graph with two kinds of edges, as shown in \textbf{Figure} \ref{fig_3}. The first one $(ld_{s,a},ld_{s,a+1}) \in \mathcal{E}_{\textbf{s}_{\Delta}}$ represents the order in an LBA delta stream $\textbf{s}_{\Delta}$ by linking $ld_{s,a}$ to its successor $ld_{s,a+1}$. The second one $(ld_{s,a},ld_{s,b}) \in \mathcal{E}_{\textbf{s}_{\Delta}}$ is built by fully connecting all nodes in the stream to capture the global information of each LBA delta stream. We denote the set of sequential edges as $\mathcal{E}^{S}_{\textbf{s}_{\Delta}}$ and full-connected edges as $\mathcal{E}^{F}_{\textbf{s}_{\Delta}}$. We compute the adjacency matrices $\mathbf{M}_S$ and $\mathbf{M}_F$ of $\mathcal{E}^{S}_{\textbf{s}_{\Delta}}$ and $\mathcal{E}^{F}_{\textbf{s}_{\Delta}}$ separately. As every LBA delta node might appear more than once in a stream, we normalize the weights on each edge in $\mathcal{E}_{\textbf{s}_{\Delta}}$. The weight of an edge is set to be its occurrence counts divided by the out-degree of its start node. The higher the weight, the stronger the correlation between the corresponding two nodes. The incoming parts of $\mathbf{M}_S$ and $\mathbf{M}_F$ are computed as Eq. (\ref{Eq_m_local}) and (\ref{Eq_m_global}),

\begin{equation}
	\textbf{M}^{in}_{S} = \sum_{i=1}^{K+1}\frac{\textbf{1}((ld_{s,a},ld_{s,a+1})_{\mathcal{E}^{S}_{\textbf{s}_{\Delta}}})}{\textbf{1}((ld_{s,i},ld_{s,a+1})_{\mathcal{E}^{S}_{\textbf{s}_{\Delta}}})}, \ (a \le n-1),
	\label{Eq_m_local}
\end{equation}
\begin{equation}
	\textbf{M}^{in}_{F} = \sum_{i=1}^{K+1}\frac{\textbf{1}((ld_{s,a},ld_{s,b})_{\mathcal{E}^{F}_{\textbf{s}_{\Delta}}})}{\textbf{1}((ld_{s,i},ld_{s,b})_{\mathcal{E}^{F}_{\textbf{s}_{\Delta}}}) \times |b-a|}, \ (a,b \le n),
	\label{Eq_m_global}
\end{equation}

\noindent where $\textbf{1}(\cdot)$ is indicator function and the outgoing parts $\textbf{M}^{out}_{F}$ and $\textbf{M}^{out}_{F}$ is computed in the same manner. Then, we expand $\textbf{M}_S$ and $\textbf{M}_F$ to the same dimension, $\textbf{M}_S, \textbf{M}_F \in \mathbb{R}^{(K+1) \times 2*(K+1)}$. To facilitate the fuse of information in $\textbf{M}_S$ and $\textbf{M}_F$, we conduct a weighted sum of them named as hybrid connection matrix $\textbf{M}_h$.

Then, the preprocessed LBA delta stream is embedded with the hybrid connection matrix $\textbf{M}_h$ and fed into a graph neural network. Specifically, we obtain each node vector $\overrightarrow{ld}_{s,i} \in \mathbb{R}^d$ that indicates the $d$-dimensional latent vector of the original $ld_{s,i} \in LD$. Each LBA delta stream $\textbf{s}_\Delta$ can be presented as an embedding matrix $\textbf{S}_{\Delta}$ by DeepWalk \cite{Perozzi2014DeepWalkOL} where each node vector $\overrightarrow{ld}_{s,i} \in \mathbb{R}^d$ denotes a $d$-dimensional real-valued latent vector. Note that SGDP can support LBA streams of various lengths and various graph model constructing strategies.

\subsection{Latent Node Vectors Updating}
We apply the vanilla graph neural network (GNN) proposed by   \cite{scarselli2008graph} and gated-recurrent-units-based GNN (gated GNN)   \cite{li2015gated} to obtain latent features of nodes. Specifically, for each LBA delta stream vector $\textbf{S}_{\Delta} = [\overrightarrow {ld}_1^{t-1},\overrightarrow {ld}_2^{t-1},...,\overrightarrow {ld}_{n}^{t-1}]$ in the graph $\mathcal{G}_{\textbf{s}_{\Delta}}$, the update functions can be formalized as,
\begin{equation}
	\mathbf{a}^t = \mathbf{A}_{\textbf{M}_h}^{t-1}(\textbf{S}_{\Delta} \mathbf{W_a}^t + \mathbf{b_a}^t),
	\label{Eq_1}
\end{equation}
\begin{equation}
	\mathbf{z}^t = \sigma(\mathbf{W_z}^t\mathbf{a}^t+\mathbf{U_z}^t\overrightarrow {ld}^{t-1}_n),
	\label{Eq_2}
\end{equation}
\begin{equation}
	\mathbf{r}^t = \sigma(\mathbf{W_r}^t\mathbf{a}^t+\mathbf{U_r}^t\overrightarrow {ld}^{t-1}_n),
	\label{Eq_3}
\end{equation}
\begin{equation}
	\tilde{\mathbf{h}}^t = tanh(\mathbf{W_h}^t\mathbf{a}^t+\mathbf{U_h}^t(\mathbf{r}^t \odot \overrightarrow {ld}^{t-1}_n),
	\label{Eq_4}
\end{equation}
\begin{equation}
	\mathbf{h}^t = (1-\mathbf{z}^t) \odot \overrightarrow {ld}^{t-1}_n + \mathbf{z}^t \odot \tilde{\mathbf{h}}^t,
	\label{Eq_5}
\end{equation}

\noindent where $\mathbf{A}_{\textbf{M}_h}^{t-1} \in \mathbb{R}^{1 \times 2n}$ is two rows of blocks (outgoing and incoming) in $\mathbf{M}_h$ corresponding to node $ld^{t-1}_n$. $\mathbf{a}^t$ extracts the contextual features of neighborhoods for node $ld^{t-1}_n$ with weight matrix $\mathbf{W_a}^t \in \mathbb{R}^{d \times 2d}$ and bias vector $\mathbf{b_a}^t \in \mathbb{R}^{2d}$. Then, we take $\mathbf{a}^t$ and previous LBA delta vector $\overrightarrow{lb}_n^{t-1}$ as input and feed them into the gated GNN. The updated functions are shown in Eq. (\ref{Eq_2})$\sim$(\ref{Eq_5}). $\mathbf{z}^t$ and $\mathbf{r}^t$ are the updates and the reset gate, and control which features to be reserved or discarded. $\sigma(\cdot)$ represents the logistic sigmoid function and $\odot$ denotes the element-wise multiplication operator. $\mathbf{W_z}^t$, $\mathbf{W_r}^t$, $\mathbf{W_h}^t$ and $\mathbf{U_z}^t$, $\mathbf{U_r}^t$, $\mathbf{U_h}^t$ are the weight matrices to be learned. The final state $\mathbf{h}^t$ is the latent node vector, which is the sum of the candidate states and the previous hidden states. Note that the model will update all nodes until they converge.

\subsection{Generating Stream Hybrid Embedding Vector}
The next accessed LBA is strongly correlated with the previous ones, and that relationship is inversely proportional to the interval between the two LBAs. Therefore, we apply a hybrid embedding vector to extract features, i.e., local embedding and global embedding. Firstly, the local embedding vector named $\mathbf{v}_l^t$ is  defined as the last accessed LBA delta $\overrightarrow {ld}_n^{t-1}$,
\begin{equation}
	\mathbf{v}_l^t = \overrightarrow{ld}_n^{t-1}.
	\label{Eq_6}  
\end{equation}

\noindent The global embedding vector  $\mathbf{v}_g^t$ aggregates all node vectors in the LBA delta stream $\textbf{S}_{\Delta}$. Specially, we use the soft-attention approach to more effectively represent the different levels of priority, as Eq. (\ref{Eq_7}) and (\ref{Eq_8}) show,
\begin{equation}
	\mathbf{\alpha}_i^t = {\mathbf{q}^t}^{\top} \sigma(\mathbf{W_1}^t \overrightarrow {ld}_n^{t-1} + \mathbf{W_2}^t \overrightarrow {ld}_i^{t-1} + \mathbf{b}_g^t),
	\label{Eq_7}
\end{equation}
\begin{equation}
	\mathbf{v}_g^t = \sum_{i=1}^n \mathbf{\alpha}_i^t \overrightarrow {ld}_i^{t-1},
	\label{Eq_8}
\end{equation}

\noindent where $\mathbf{W_1}^t,\mathbf{W_2}^t \in \mathbb{R}^{d \times d}$ and $\mathbf{q}^t \in \mathbb{R}^d$ are weight matrices, and $\mathbf{b}_g^t \in \mathbb{R}^d$ is bias vector.

Finally, the hybrid embedding vector $\mathbf{v}_h^t$ combines the local embedding vector $\mathbf{v}_l^t$ and the global embedding vector $\mathbf{v}_g^t$ linearly, as the Eq. (\ref{Eq_9}) shows,
\begin{equation}
	\mathbf{v}_h^t = \mathbf{W_f}^t[\mathbf{v}_l^t;\mathbf{v}_g^t] + \mathbf{b_h}^t,
	\label{Eq_9}
\end{equation}

\noindent where $\mathbf{W_f}^t \in \mathbb{R}^{d \times 2d}$ is weight matrix and $\mathbf{b_h}^t \in \mathbb{R}^d$ is bias vector. The final hybrid embedding vector $\mathbf{v}_h^t$ of the LBA delta stream $\textbf{S}_{\Delta}$ is in the $d$ dimensions.

\subsection{Forecasting and Prefetching}\label{algrchapt}
After extracting the hybrid embedding vector $\mathbf{v}_h^t$, we predict the score of each LBA delta candidate $\widehat{\mathbf{z}}_i^t$  in stream $\textbf{s}_{\Delta}$ by multiplying $\mathbf{v}_h^t$ and $\overrightarrow {ld}_i^{t-1}$ as
\begin{equation}
	\widehat{\mathbf{z}}_i^t = {\mathbf{v}_h^t}^{\top} \overrightarrow{ld}_i^{t-1}.
	\label{Eq_10}
\end{equation}

We take the candidate with the highest score as the predicted LBA delta. Besides, in order to train with labels, we need to compute the probability of each node $\widehat{\mathbf{y}}^t \in \mathbb{R}^{K+1}$ in the next step using the softmax function, which is
\begin{equation}
	\widehat{\mathbf{y}}^t = \mathit{Softmax}(\widehat{\mathbf{z}}^t).
	\label{Eq_11}
\end{equation}

The loss function is the cross-entropy of prediction $\widehat{\mathbf{y}}^t$ and ground truth $\mathbf{y}^t$ with regularization, as shown in Eq. (\ref{Eq_12}).
\begin{equation}
	\mathcal{L}(\widehat{\mathbf{y}}^t) = -\sum_{i=1}^m [\mathbf{y}_i^t \mathit{log}(\widehat{\mathbf{y}}_i^t) + (1-\mathbf{y}_i^t)\mathit{log}(1-\widehat{\mathbf{y}}_i^t)] + \lambda \lVert \theta\rVert^2_2,
	\label{Eq_12}
\end{equation}
where $\lambda$ is $\mathit{l2}$-norm penalty factor, $\theta$ is weight vectors. 

Finally, we decode the index of $\max \widehat{\mathbf{z}}_i^t$ to $\widehat{ld}_{s,n+1}$, predict the next accessed LBA by Eq. (\ref{Eq_0}), and prefetch the corresponding block from storage into the cache. Overall, we summarize the proposed SGDP framework in \textbf{Algorithm 1}.

\begin{algorithm}[!h]
	\renewcommand{\arraystretch}{1}
	\renewcommand{\baselinestretch}{1}
    \caption{The Workflow of \textbf{SGDP} Framework}
    \label{SGDPalg}
    {\bf Input:} An LBA sequence $\langle lba_i \rangle_{i=1}^{l} = \langle lba_1,lba_2,\ldots,lba_l \rangle$ , the top number of most frequent LBA delta $K$, dimension of embedding vector of each LBA delta stream $d$, cache size $N$, maximum iteration $Q$, stop criteria $tol$.
    \begin{algorithmic}[1]
    \STATE \textbf{\underline{Step 1: LBA stream preprocessing}}
    \STATE Compute delta of each adjacent LBA pair in $\langle lba_i \rangle_{i=1}^{l}$ and get LBA delta $\langle ld_i \rangle_{i=1}^{l-1} = \langle ld_1,ld_2,\ldots,ld_{n-1} \rangle$.
    \STATE Encode $ld$s in all LBA streams by $top(K)$ to $LD$.
    \STATE Generate LBA delta stream $\textbf{s}_{\Delta}$ by time limit and slide window.
    \STATE \textbf{\underline{Step 2: Embed LBA delta stream to a graph}}
    \STATE Compute $\textbf{M}_S$ by Eq. (\ref{Eq_m_local}) and $\textbf{M}_F$ by Eq. (\ref{Eq_m_global}).
    \STATE Expand $\textbf{M}_S$ and $\textbf{M}_F$ and conduct weighted sum to get $\textbf{M}_h$.
    \STATE Conduct embedding of each $\textbf{s}_\Delta \in \mathbb{R}^{n}$ into $\textbf{S}_\Delta \in \mathbb{R}^{d \times n}$.
    \STATE \textbf{\underline{Step 3: Update hybrid embedding vector and training}}
   \STATE Initialize the parameters in the gated GNN model.
   \FOR{$q = 1,..., Q$}
    \STATE Compute the Eq.(\ref{Eq_1}) $\sim$ Eq.(\ref{Eq_12}) to fit each stream with the input $\textbf{M}_h$ and $\textbf{S}_\Delta$.
   \STATE Update the weight matrices list \{$\mathbf{W_z}$, $\mathbf{W_r}$, $\mathbf{W_h}$, $\mathbf{U_z}$, $\mathbf{U_r}$, $\mathbf{U_h}$, $\mathbf{W_1}$, $\mathbf{W_2}$, $\mathbf{W_f}$\} and vectors list \{$\mathbf{b_g}$, $\mathbf{b_h}$\} by Adam with ground truth $\textbf{y}$.
    \STATE Convergence checking: if $\mathcal{L}(\widehat{\mathbf{y}}) < tol$, break; otherwise, continue.
    \ENDFOR
    \STATE \textbf{\underline{Step 4: Conduct forecasting and data prefetching}}
    \STATE Conduct Eq.(\ref{Eq_10}) and get $\max \widehat{\textbf{z}}$ with the updated model.
    \STATE Decode it to $\widehat{ld}_{n}$ and conduct Eq.(\ref{Eq_0}) to get $\widehat{lba}_{n+1}$.
    \STATE Read the corresponding block and prefetch it into the cache or conduct no prefetching.
    \end{algorithmic}
    {\bf Output:} $\widehat{lba}_{n+1}$, blocks $ \in \mathbb{R}^N$ in cache.
\end{algorithm}

\section{Experimental Settings}

\subsection{Datasets}
We use representative eight datasets in production servers from different applications, including six datasets from an open-source benchmark \textbf{MSRC} and two datasets from a real enterprise storage system \textbf{HW}:

\textbf{MSRC}\footnote{http://iotta.snia.org/traces/388} (Microsoft Research Cambridge) \cite{lee2017understanding}: It collects a 1-week LBA sequence of live enterprise servers at Microsoft. We use its five datasets from different application scenes named \{\textbf{hm\_1}, \textbf{mds\_0}, \textbf{proj\_0}, \textbf{prxy\_0}, \textbf{src1\_2}\}. 

\textbf{HW}: It consists of three datasets collected from a real-world commercial hybrid storage system, which describes storage traffic characteristics on enterprise virtual desktop infrastructure and production servers. It intercepts the stream from an intra-enterprise storage system under different application scenarios and reads the storage system record logs directly. We named these three datasets as \{\textbf{hw\_1}, \textbf{hw\_2}, \textbf{hw\_3}\}.

\textbf{Table} \ref{tab:Datasets-description} provides the detail of the eight datasets from two data sources. Memory means the total amount of storage space that has been accessed in the trace. Sequential shows the percentage of sequential accesses in the trace.

\begin{table}[!t]
	\caption{Datasets Description}
	\centering
	\renewcommand{\arraystretch}{1.0}
	\resizebox{0.49\textwidth}{!}{
	\begin{tabular}{cccccc}
    	\hline \hline
        Source & Dataset & Length &  Memory (GB) & Function  &Sequential (\%)\\ \midrule
        \multirow{5}*{MSRC} & hm\_1 & 1.08$\times10^6$ & 6.36 & Hardware monitoring &39.9\\
         & mds\_0 & 4.23$\times10^5$ & 8.48 & Media server & 65.2\\
         & proj\_0 & 1.17$\times10^6$ & 4.056 & Project directories  & 57.3
\\
         & prxy\_0 & 4.03$\times10^5$ & 5.18 & Firewall/web proxy &37.6
\\
         & src1\_2 & 1.15$\times10^6$ &  2.0 & Source control  &58.5
\\
  \hline      
        \multirow{3}*{HW} & hw\_1 & 1.39$\times10^6$ &  930.29 &  hybrid storage system & 55.8
\\
         & hw\_2 & 2.58$\times10^5$ &  600.46 &  hybrid storage system &95.1
\\
         & hw\_3 & 1.73$\times10^5$ & 902.22 &  hybrid storage system &43.7
\\
		\hline \hline
	\end{tabular}}
	\label{tab:Datasets-description}
\end{table}

\subsection{Compared Methods}
We compare SGDP with the following methods from three categories: traditional prefetchers, regression-based prefetchers, and learning-based prefetchers.

\textbf{No\_pre} means without any prefetching facilities and is used as a baseline to show the gain by other schemes.

\textbf{Naïve Prefetcher} treats the LBA stream as a whole sequence, i.e., $\widehat{ld}_{n} = ld_{n-1}$, and directly uses Eq. (\ref{Eq_0}) to predict LBA.

\textbf{Stride Prefetcher} \cite{ki2000stride} simultaneously records 128 LBA access streams, and each of them tracks the last 3 LBA accesses. Each access is mapped to a stream based on hashing the most significant LBA. If the difference between the 3 LBA accesses matches, it will detect a stride and conduct a prediction.

\textbf{ARIMA}\cite{tran2004automatic} treats the problem as a time-series prediction problem and applies the ARIMA model built from $t-\delta$ to $t$ to forecast the next LBA.

\textbf{Informer\footnote{https://github.com/zhouhaoyi/Informer2020/}}   \cite{zhou2021informer} is the SOTA for time-series forecasting. Same as ARIMA, it takes the previous LBA delta sequence as input and predicts the following LBA.

\textbf{DeepPrefetcher}   \cite{ganfure2020deepprefetcher} captures the LBA delta patterns by employing the word2vec model and
LSTM architecture for prefetching.  

\textbf{Delta-LSTM\footnote{https://github.com/Chandranil2606/Learning-IO-Access-Patterns-to-improve-prefetching-in-SSDs-/}} \cite{chakraborttii2020learning} is another learning-based prefetcher, which uses an LSTM-based model to predict the LBA delta for prefetching. 

Besides, we set the sequences of LBA delta as the input for a fair comparison. There is a class explosion problem in DeepPrefetcher, which makes it unrealistic to train. To solve the problem, we restrict the top-$K$ class with $K=10000$. This $K=10000$ value is to balance the efficiency and accuracy based on our preliminary study. For Delta-LSTM and SGDP, we set top-$1000$ frequently occurring LBA deltas as input.

\subsection{Evaluation Criteria}

\textbf{HR@N} (Hit Ratio) is the number of cache hits divided by the total number of memory requests over a given time interval. It is an important storage indicator given a fixed cache size $N$. 
That is,	
\begin{equation}
	\rm HR = \frac{\rm Cache \, \,Hits}{\rm Cache \,Hits + Cache \, \,Misses} \times 100 \%. 
\end{equation}

\textbf{EPR@N} (Effective Prefetching Ratio) is the ratio of the number of correctly prefetched data to all executed prefetches given a fixed cache size $N$, which is strongly related to the prefetcher's efficiency. 
That is, 
\begin{equation}
	\rm EPR = \frac{\rm Correct \, \,Prefetchings}{\rm All \, \,Prefetchings} \times 100 \%.
\end{equation}

Note that HR and EPR describe the prefetching results more precisely and feasible refer to Accuracy and Recall in DeepPrefetcher \cite{ganfure2020deepprefetcher} and Delta-LSTM \cite{chakraborttii2020learning}, respectively. We use HR and EPR in our work because they describe the prefetching results more precisely. Besides, there is a trade-off between HR and EPR in the subsection \ref{SGDP1Varints} and some results are shown in \textbf{Figure} \ref{fig_trade}.

Besides, we feed the next step predicted LBA into the cache simulator based on the Least Recently Used (LRU) strategy for prefetching. LRU is a classical cache elimination algorithm. It selects the most recently unused LBA to retire.

\subsection{Implementation Details}
We implemented SGDP for offline training and online testing by PyTorch \cite{paszke2017automatic}. All experiments are trained and tested on a computing server equipped with an Intel Xeon Platinum 8180M CPU@2.50GHz and an NVIDIA Tesla V100 GPU.

For a more fair and effective comparison, we normalize all LBA in increments of 8KB blocks and according to the I/O size of the 8KB block alignment and increment operations. We apply the 10-fold cross-validation method for training and testing, the same as SOTA methods. As for neural networks, all parameters are initialized with a Gaussian distribution with a means of 0 and a standard deviation of 0.1 with the latent vectors $d$ equaling 200 for all 8 datasets. Moreover, we set the initial learning rate to 1.5$\times10^{-3}$ with decay by 95\% after every 3 epochs, the batch size to 128 with 10 epochs, and the $\mathit{l2}$-norm penalty factor to 10$^{-5}$. Adam optimizer \cite{kingma2014adam} with default parameter is applied for optimization. We set the stream split time interval $T$ as 0.1ms for HW and 0.01ms for MSRC, and set the top number of most frequent LBA delta $K$ as 1000 for SGDP. Note that since the preprocessed LBA stream is shorter, the training epoch can be smaller to prevent over-fitting, which is also useful to shorten the training time.

\begin{table*}[!t]
   \centering
   \caption{Single-step Results. The results are in percentage, the best results are in \textbf{bold}, the second ones are \underline{underlined}, $N$ is the cache size.}
   \renewcommand{\arraystretch}{0.863}
   \resizebox{\textwidth}{!}{
\begin{tabular}{r|ccc|ccc|ccc|ccc|ccc|ccc|ccc|ccc}
\hline \hline
Dataset & \multicolumn{6}{c|}{\textbf{hw\_1}}           & \multicolumn{6}{c|}{\textbf{hw\_2}}           & \multicolumn{6}{c|}{\textbf{hw\_3}}           & \multicolumn{6}{c}{\textbf{hm\_1}} \bigstrut\\
\hline
\multicolumn{1}{c|}{\multirow{2}[4]{*}{\diagbox{Method}{Metric}}} & \multicolumn{3}{c|}{HR@N} & \multicolumn{3}{c|}{EPR@N} & \multicolumn{3}{c|}{HR@N} & \multicolumn{3}{c|}{EPR@N} & \multicolumn{3}{c|}{HR@N} & \multicolumn{3}{c|}{EPR@N} & \multicolumn{3}{c|}{HR@N} & \multicolumn{3}{c}{EPR@N} \bigstrut\\
\cline{2-25}      & 10    & 100   & 1000  & 10    & 100   & 1000  & 10    & 100   & 1000  & 10    & 100   & 1000  & 10    & 100   & 1000  & 10    & 100   & 1000  & 10    & 100   & 1000  & 10    & 100   & 1000 \bigstrut\\
\hline
No\_pre & 0.0   & 0.3   & 54.2  & 0.0   & 0.0   & 0.0   & 1.0   & 1.1   & 1.1   & 0.0   & 0.0   & 0.0   & 0.0   & 0.1   & 1.3   & 0.0   & 0.0   & 0.0   & 2.7   & 25.3  & 98.3  & 0.0   & 0.0   & 0.0  \bigstrut[t]\\
Naïve & 57.5  & 58.0  & 63.2  & 63.3  & 64.5  & 64.5  & 92.5  & 92.6  & 92.7  & 93.3  & 93.7  & 94.0  & 47.7  & 47.9  & 48.8  & 48.0  & 48.3  & 48.7  & 31.7  & 43.8  & 97.4  & 30.5  & 31.2  & 5.6  \\
Stride & 43.7  & 44.0  & 65.8  & 80.5  & 81.1  & 80.6  & 91.0  & 91.1  & 91.1  & \textbf{99.1} & \textbf{99.2} & \textbf{99.2} & 38.4  & 38.6  & 39.6  & 81.6  & 82.0  & 82.3  & 27.1  & 47.0  & 99.1  & \underline{82.3} & \underline{84.4} & \textbf{88.4} \\
ARIMA & 1.9   & 4.0   & 8.8   & 1.9   & 4.3   & 6.2   & 82.8  & 82.9  & 83.0  & 85.9  & 86.2  & 86.4  & 0.3   & 0.3   & 1.3   & 0.2   & 0.3   & 0.3   & 3.5   & 19.0  & 95.2  & 2.7   & 5.2   & 2.5  \\
Informer & 0.2   & 0.9   & 5.8   & 0.3   & 0.9   & 2.9   & 1.0   & 1.1   & 1.1   & 0.0   & 0.0   & 0.0   & 0.0   & 0.0   & 0.9   & 0.0   & 0.0   & 0.0   & 1.1   & 14.0  & 90.4  & 0.1   & 0.7   & 0.7  \\
DeepPrefetcher & 74.3  & 74.6  & 79.2  & 75.4  & 75.9  & 76.5  & 92.2  & 92.5  & 92.8  & 93.4  & 94.0  & 94.5  & 50.4  & 50.7  & 51.7  & 50.4  & 50.7  & 51.2  & 38.5  & 59.1  & 99.3  & 38.5  & 56.0  & 46.1  \\
Delta-LSTM & 74.4  & 74.8  & 79.3  & 75.5  & 76.0  & 76.6  & 92.5  & 92.8  & 93.1  & 93.7  & 94.2  & 94.7  & 56.4  & 56.8  & 57.9  & 66.2  & 66.8  & 67.2  & 30.0  & 50.6  & 99.3  & 57.7  & 72.8  & 87.6  \bigstrut[b]\\
\hline
SGDP  & \textbf{79.2} & \textbf{79.5} & \textbf{85.8} & \textbf{82.9} & \textbf{83.5} & \textbf{81.6} & \underline{93.0} & 93.0  & 93.1  & \underline{97.5} & \underline{97.7} & \underline{97.8} & \underline{76.0} & \underline{76.6} & \underline{77.5} & \textbf{88.9} & \textbf{89.5} & \textbf{90.1} & 38.1  & 55.7  & \textbf{99.4} & \textbf{87.8} & \textbf{90.1} & \underline{86.2} \bigstrut[t]\\
SGDP$_{l}$ & \underline{78.5} & \underline{78.8} & \underline{84.9} & \underline{82.1} & \underline{82.7} & \underline{80.6} & 92.9  & \underline{93.1} & \underline{93.2} & 97.0  & 97.2  & 97.4  & \textbf{78.5} & \textbf{79.0} & \textbf{79.8} & \underline{83.6} & \underline{84.2} & \underline{84.7} & \underline{43.1} & \underline{61.4} & 99.1  & 46.3  & 60.8  & 24.4  \\
SGDP$_{p}$ & 75.7  & 78.2  & 83.6  & 77.6  & 80.4  & 79.6  & \textbf{93.7} & \textbf{94.0} & \textbf{94.2} & 94.4  & 95.0  & 95.4  & 48.1  & 48.3  & 49.6  & 72.1  & 73.1  & 75.1  & \textbf{43.9} & \textbf{62.9} & \underline{99.4} & 46.8  & 63.8  & 34.8  \bigstrut[b]\\
\hline
Dataset & \multicolumn{6}{c|}{\textbf{mds\_0}}          & \multicolumn{6}{c|}{\textbf{proj\_0}}         & \multicolumn{6}{c|}{\textbf{prxy\_0}}         & \multicolumn{6}{c}{\textbf{src1\_2}} \bigstrut\\
\hline
\multicolumn{1}{c|}{\multirow{2}[4]{*}{\diagbox{Method}{Metric}}} & \multicolumn{3}{c|}{HR@N} & \multicolumn{3}{c|}{EPR@N} & \multicolumn{3}{c|}{HR@N} & \multicolumn{3}{c|}{EPR@N} & \multicolumn{3}{c|}{HR@N} & \multicolumn{3}{c|}{EPR@N} & \multicolumn{3}{c|}{HR@N} & \multicolumn{3}{c}{EPR@N} \bigstrut\\
\cline{2-25}      & 10    & 100   & 1000  & 10    & 100   & 1000  & 10    & 100   & 1000  & 10    & 100   & 1000  & 10    & 100   & 1000  & 10    & 100   & 1000  & 10    & 100   & 1000  & 10    & 100   & 1000 \bigstrut\\
\hline
No\_pre & 13.2  & 35.0  & 61.0  & 0.0   & 0.0   & 0.0   & 6.1   & 28.7  & 35.2  & 0.0   & 0.0   & 0.0   & 20.1  & 40.7  & 48.8  & 0.0   & 0.0   & 0.0   & 3.9   & 34.8  & 48.2  & 0.0   & 0.0   & 0.0  \bigstrut[t]\\
Naïve & 54.3  & 68.2  & 85.2  & 47.8  & 51.1  & 52.2  & 61.1  & 70.1  & 74.3  & 58.7  & 59.7  & 60.8  & 46.4  & 64.3  & 72.7  & 35.1  & 38.5  & 40.9  & 60.5  & 73.0  & 80.8  & 59.9  & 63.1  & 66.3  \\
Stride & 47.3  & 62.2  & 79.8  & \textbf{82.3} & \textbf{90.6} & \textbf{89.8} & 51.0  & 61.1  & 65.4  & 82.5  & \textbf{88.1} & \textbf{88.3} & 40.3  & 56.5  & 63.8  & 69.6  & \underline{81.1} & \underline{81.4} & 48.3  & 63.8  & 73.4  & 81.0  & \textbf{89.6} & \textbf{92.0} \\
ARIMA & 16.6  & 37.4  & 58.3  & 8.6   & 9.2   & 12.0  & 12.9  & 33.5  & 39.3  & 12.0  & 10.1  & 10.4  & 19.9  & 42.2  & 52.3  & 6.5   & 7.2   & 8.2   & 14.6  & 42.0  & 54.8  & 19.5  & 17.7  & 19.2  \\
Informer & 9.6   & 28.3  & 54.5  & 0.3   & 1.2   & 5.2   & 3.9   & 19.8  & 34.7  & 0.1   & 0.5   & 2.3   & 13.7  & 32.1  & 46.9  & 0.0   & 0.0   & 0.2   & 1.7   & 22.5  & 45.3  & 0.0   & 0.1   & 0.6  \\
DeepPrefetcher & 60.7  & 73.7  & 88.5  & 66.9  & 77.5  & 83.3  & 72.6  & 79.1  & 82.8  & 75.0  & 78.6  & 81.5  & 57.0  & 70.2  & 77.4  & 63.5  & 70.4  & 73.9  & 74.5  & 82.9  & 89.0  & 76.2  & 80.9  & 87.0  \\
Delta-LSTM & 57.3  & 69.6  & 86.2  & 80.2  & \underline{87.8} & \underline{89.8} & 62.3  & 69.1  & 73.3  & \textbf{84.3} & 86.2  & 87.4  & 52.2  & 64.2  & 71.3  & \underline{75.7} & 79.3  & 80.9  & 70.0  & 79.6  & 86.2  & 77.5  & 81.3  & 87.2  \bigstrut[b]\\
\hline
SGDP  & 66.0  & 76.3  & 91.6  & \underline{80.2} & 87.0  & 88.4  & 73.4  & 78.5  & 82.1  & \underline{84.0} & \underline{87.6} & \underline{88.2} & \underline{62.2} & 73.2  & 79.9  & \textbf{76.3} & \textbf{83.3} & \textbf{84.1} & \underline{75.4} & 83.1  & 88.8  & \textbf{82.5} & \underline{88.5} & \underline{90.8} \bigstrut[t]\\
SGDP$_{l}$ & \underline{66.1} & \underline{77.5} & \underline{92.1} & 65.4  & 73.9  & 79.6  & \textbf{75.5} & \underline{81.1} & \underline{84.6} & 79.8  & 83.6  & 85.6  & \textbf{64.1} & \textbf{76.5} & \underline{83.0} & 65.7  & 74.3  & 78.4  & \textbf{76.3} & \underline{83.9} & \underline{89.4} & \underline{81.5} & 87.4  & 89.3  \\
SGDP$_{p}$ & \textbf{67.4} & \textbf{79.8} & \textbf{92.6} & 68.7  & 81.9  & 87.9  & \underline{73.7} & \textbf{81.3} & \textbf{85.2} & 74.5  & 80.2  & 83.9  & 63.9  & \underline{76.2} & \textbf{83.0} & 64.9  & 72.6  & 76.1  & 74.9  & \textbf{84.8} & \textbf{89.8} & 76.1  & 84.0  & 87.3  \bigstrut[b]\\
\hline \hline
\end{tabular}%
}
\label{tab:addlabel}
\end{table*}%

\begin{table*}[!t]
   \centering
   \caption{Average Results of Multi-step Prefetching. The results are in percentage, the best results are in \textbf{bold}, and the cache size is 100.}
   \renewcommand{\arraystretch}{1.0}
\resizebox{\textwidth}{!}{

\begin{tabular}{rcccccccccc|cccccccccc}
\hline \hline
\multicolumn{1}{r|}{Metric} & \multicolumn{10}{c|}{HR@100}   & \multicolumn{10}{c}{EPR@100} \bigstrut\\
\hline 
\multicolumn{1}{l|}{\diagbox{Method}{Step}} & 1     & 2     & 3     & 4     & 5     & 6     & 7     & 8     & 9     & 10    & 1     & 2     & 3     & 4     & 5     & 6     & 7     & 8     & 9     & 10 \bigstrut\\
\hline
\multicolumn{1}{r|}{DeepPrefetcher} & 74.0  & 76.1  & 76.7  & 77.1  & 77.2  & 77.1  & 77.1  & 77.0  & 77.0  & 76.9  & 73.1  & 60.7  & 52.5  & 46.5  & 42.0  & 38.3  & 35.3  & 32.8  & 30.7  & 28.8  \bigstrut[t]\\
\multicolumn{1}{r|}{Delta-LSTM} & 70.2  & 74.9  & 77.1  & 78.0  & 78.5  & 78.9  & 79.3  & 79.6  & 79.8  & 80.0  & 81.5  & 72.3  & 65.9  & 60.6  & 56.3  & 52.6  & 49.5  & 46.8  & 44.5  & 42.3  \bigstrut[b]\\
\hline
\multicolumn{1}{r|}{SGDP}  & 77.0  & 78.3  & 78.9  & 79.3  & 79.6  & 79.8  & 80.0  & 80.1  & 80.3  & 80.3  & \textbf{88.4} & \textbf{80.5} & \textbf{74.0} & \textbf{68.9} & \textbf{64.6} & \textbf{60.9} & \textbf{57.7} & \textbf{54.9} & \textbf{52.7} & \textbf{50.5} \bigstrut[t]\\
\multicolumn{1}{r|}{SGDP$_{l}$} & \textbf{78.9} & \textbf{80.7} & \textbf{81.4} & \textbf{81.8} & \textbf{82.2} & \textbf{82.4} & \textbf{82.5} & \textbf{82.6} & \textbf{82.7} & \textbf{82.7} & 80.5  & 70.2  & 62.9  & 57.5  & 53.1  & 49.6  & 46.6  & 44.0  & 41.8  & 39.8  \\
\multicolumn{1}{r|}{SGDP$_{p}$} & 75.7  & 77.4  & 78.1  & 78.6  & 78.9  & 79.1  & 79.2  & 79.3  & 79.5  & 79.5  & 78.9  & 68.3  & 60.8  & 55.1  & 50.6  & 46.8  & 43.7  & 41.1  & 38.8  & 36.7 \\
\hline  \hline
\label{roll_results} 
\end{tabular}%
}
\end{table*}

\section{Experimental Results}
\subsection{Results of Single-Step Prefetching}
We analyze the results of data prefetching conducted by  SGDP and the compared methods on different datasets in the case of single-step Prefetching, as reported in \textbf{Table} \ref{tab:addlabel}. The detailed analysis is presented in the following.

\textbf{ARIMA and Informer} Time-series forecasting models (ARIMA and Informer) perform the worst, excepting the case of hw\_2 trace where ARIMA achieves around 80\% HR as this dataset has a very high degree of sequential access (95.1\%). ARIMA and Informer take LBA delta inputs as scalar variables and can produce a correct prediction if the input sequences are steady. As there are frequent large fluctuations in complex non-sequential access patterns, ARIMA and Informer prompt incorrect LBA access. The worst results shown in almost all cases confirm our claims that regression-based approaches are not feasible for accurate and complex data prefetching.  

\textbf{Naïve and Stride Prefetcher} Traditional prefetchers (Naïve and Stride) have relatively stable performances in sequential access. However, for the random accesses, those traditional prefetchers encounter a big gap compared to SGDP. Moreover, Stride always achieves higher EPR while lower HR as it is more laziness and only prefetches when detecting an inside-page stride. As a result, it prefetches less and has higher accuracy for sequential access. In other words, although Stride has a high EPR, it prefetches less and gets quite low HR, which makes it impractical. Overall, the robustness performance of the Naïve and Stride Prefetcher is poor, especially for completely random access.

\textbf{Delta-LSTM, DeepPrefetcher and SGDP} Learning-based prefetchers (Delta-LSTM, DeepPrefetcher and SGDP) cover all highest HR and almost the highest EPR. SGDP has higher HR than DeepPrefetcher/Delta-LSTM in 19/24 out of all 24 cases, and 24/19 about EPR. Specifically, Delta-LSTM and DeepPrefetcher share a similar structure and show a similar effect. DeepPrefetcher has a larger LBA delta candidates pool which leads to more prefetching and lowers accuracy, reflected in lower average EPR (71.6\% compared to 80.3\%). On the contrary, Delta-LSTM prefetches more accurately but with fewer blocks, which leads to lower HR (70.8\% compared to 73.5\%). To be fair, as SGDP takes the top-1000 LBA delta as input (same as Delta-LSTM), the comparison to Delta-LSTM can prove that SGDP has better feature extraction ability.

\subsection{Ablation study and SGDP Variants by Stream Construction} 

\subsubsection{Retaining top-$K$ delta value (SGDP$_{l}$)}  \label{SGDP1Varints}
SGDP considers the top-1000 most frequently occurring LBA delta in the whole search space. Considering more LBA delta values further increases the model coverage, but also increases the search space and degrades the accuracy of the model. So to explore it quantitatively and prove the HR-EPR trade-off, we apply the top-10000 LBA delta for model building and name it SGDP$_{large}$, or \underline{SGDP$_{l}$} for brevity.


The experiment of SGDP$_{l}$ confirms the trade-off of HR-EPR. As \textbf{Table} \ref{tab:addlabel} shows, SGDP$_{l}$ achieves  1.5\%  slightly higher than SGDP in terms of HR while showing a large gap  with 9.6\% loss than SGDP in terms of EPR. Compared to DeepPrefetcher, which also uses top-10000 LBA delta as input, SGDP$_{l}$ maintains higher HR in 23 cases and higher EPR in 20 cases. Notice that for dataset hm\_1, SGDP$_{l}$ gets a low EPR@1000 (24.4\%). The reason is that with an extremely high HR@1000 (99.4\%, almost all hit), SGDP$_{l}$ prefetches extra useless blocks without harm to HR, leading to an obvious decline of EPR.

\begin{figure}[!t]
	\centering
	\includegraphics[width=0.8\columnwidth]{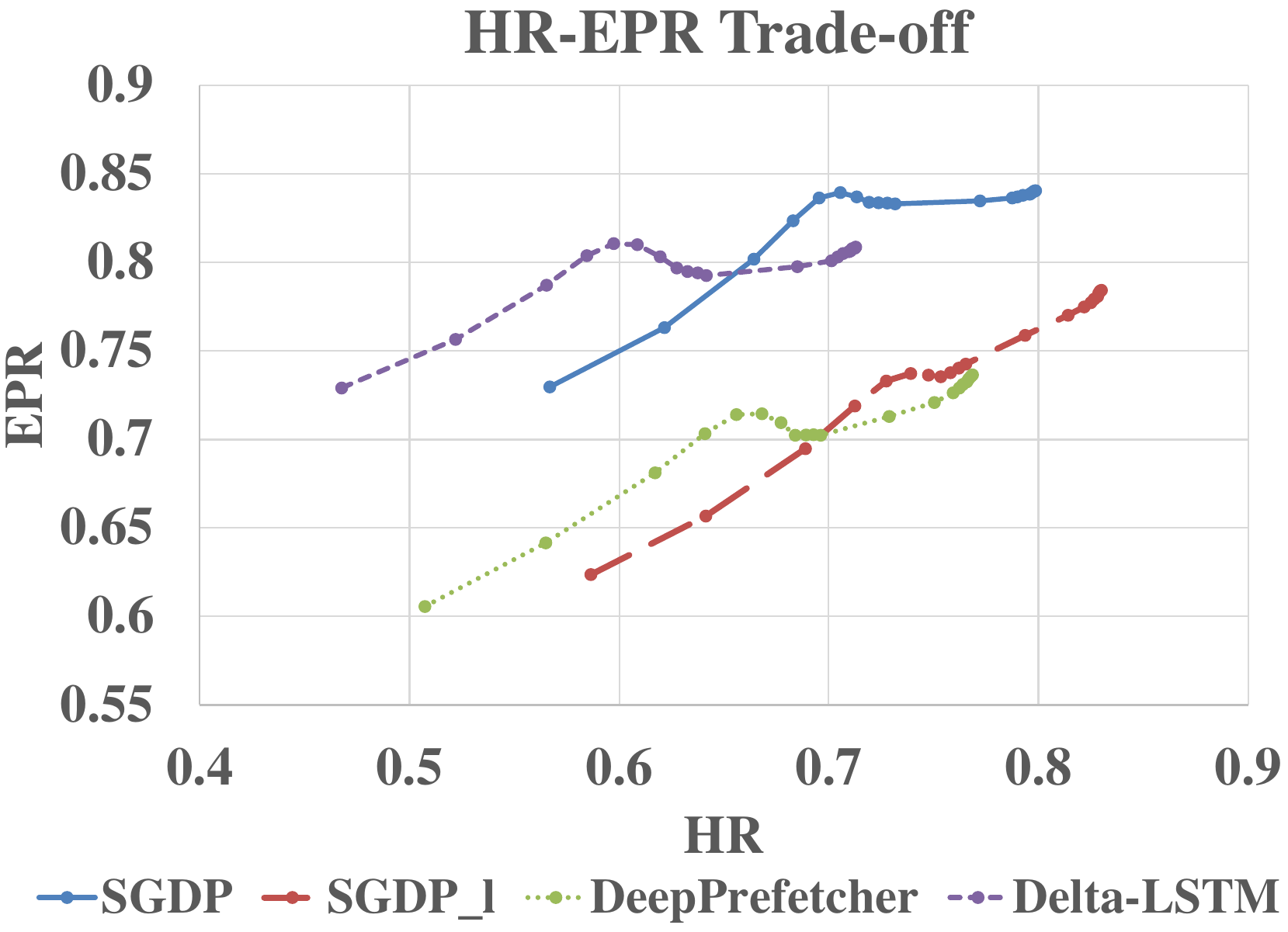}
	\caption{HR-EPR trade-off.}
	\label{fig_trade}
\end{figure}

To further verify the effectiveness of SGDP and the HR-EPR trade-off, we conduct extra tests on dataset prxy\_0 by testing Delta-LSTM, DeepPrefetcher, SGDP and SGDP$_{l}$ on 20 different cache sizes (\{5, 10, 20, $\cdots$ ,90 and 100, 200, $\cdots$, 900, 1000\}). As shown in \textbf{Figure} \ref{fig_trade}, the top-10000 methods (SGDP$_{l}$ and DeepPrefetcher) show higher HR but lower EPR than their top-1000 counterparts (SGDP and Delta-LSTM). The two pairs of top-$K$ comparisons confirm that SGDPs achieve better performance consistently than other methods.

\subsubsection{Stream partition with page (SGDP$_{p}$)}

Considering the spatially localized relevance of LBA access patterns, we divide the entire search space by 64MB size page, record the LBA access streams on each page simultaneously, and perform parallel prediction in each page stream for prefetching the next block inside the page stream. We call it \underline{SGDP$_{p}$}. SGDP$_{p}$ models LBA deltas, as 64MB page contains 8192 unique blocks, the total LBA delta candidate class of SGDP$_{p}$ is 16383 ($\pm 8191$) instead of top-$K$ classes.

SGDP$_{p}$ keeps the best HR results on half of all cases. Its average HR is slightly lower than SGDP and SGDP$_{l}$, but still higher than other methods. HR of SGDP$_{p}$ on hw\_3 encounters a severe drop. The reason is that hw\_3 is the shortest and has the second-largest storage capacity, which means the LBAs are much more sparse. SGDP$_{p}$ needs at least an LBA stream with length 2 to generate an LBA delta stream. But as hw\_3 cross over $1.4\times10^{4}$ pages, the average length of inside-page-stream is 12.4, which means SGDP could not perform prefetching on 1/12 of data points. Nevertheless, SGDP$_{p}$ obtains the highest HR (79.9\%) on all datasets except hw\_3 on average. Therefore, it could be concluded that SGDP$_{p}$ performs well in real-world scenarios with sufficient data.


\subsection{Multi-step Prefetching}

We further evaluate the performance of SGDP methods in multi-step prefetching based on rolling prediction. We feed the prediction of LBA back to the aforementioned learning-based prefetchers and get the rolling prediction for the next LBAs. The experiments are performed on cache size 100 with a rolling step from 2 to 10. The average results of HR@100 and EPR@100 are reported in \textbf{Table} \ref{roll_results}. Overall, SGDP$_{l}$ has the best HR@100 on all steps on average, and SGDP achieves the best EPR@100. SGDP$_{p}$ shows worse results than SGDP and SGDP$_{l}$. SGDP$_{p}$ gets the highest HR@100 (from 79.6\% to 83.8\%) on average in all steps on all the datasets except hw\_3  as it is too sparse. The second best method is SGDP$_{l}$, of which HR@100s range from 78.9\% to 83.0\%. These results demonstrate that SGDP methods are able to keep their superiority and robustness in multi-step prefetching.


\begin{table}[!t]
	\centering
	\caption{The number of predictions inferred per second by learning-based methods.}
	\renewcommand{\arraystretch}{1}
	\resizebox{0.98\columnwidth}{!}{
        \begin{tabular}{r|rrrrrrrr|c}
        \hline  \hline
        \diagbox{Method}{Dataset} & hw\_1  & hw\_2  & hw\_3  & hm\_1  & mds\_0 & proj\_0 & prxy\_0 & src1\_2 & avg    \\ \hline 
        Delta-LSTM & 89.4 & 87.4 & 94.5 & 92.4 & 90.7 & 91.5 & 88.4 & 95.1 & 91.2 \\
        DeepPrefetcher & 208.2 & 154.5 & 194.2 & 160.1 & 248.4 & 178.4 & 187.9 & 249.6 & 197.7 \\ \hline 
        SGDP & \textbf{644.5} & \textbf{692.4} & \textbf{666.1} & 515.2 & 543.5 & 553.3 & 470.0 & 550.7 & 579.5 \\
        SGDP$_{l}$ & 634.7 & 686.9 & 614.7 & 500.1 & \textbf{651.4} & \textbf{663.9} & 526.3 & \textbf{670.7} & \textbf{618.6} \\
        SGDP$_{p}$ & 599.5 & 645.6 & 593.9 & \textbf{567.0} & 491.7 & 529.3 & \textbf{574.8} & 558.7 & 570.1 \\ \hline \hline
        \end{tabular}
        }
	\label{tab:efficiency}
\end{table}

\subsection{Offline Training and Online Testing Efficiency}


The offline training time in dataset hw\_1 for SGDPs is 0.37 hours, and the training time for other learning-based methods is about 1.1 hours. The GPU utilization is 24\%, the parameter number is 192,500, and the flop number is 48,570,779. For the online inference test, the GPU utilization is 10\%, the model size is 1.47 MB, and the flop number is 1,884,055. Furthermore, to verify the practicality of SGDP compared to the other methods, we collect statistics of inference time as shown in \textbf{Table} \ref{tab:efficiency}. SGDPs process 469 to 670 LBA deltas per second, while Delta-LSTM and DeepPrefetcher can only process 91.2 and 197.7 on average. SGDP$_{l}$ speeds up inference time up to 3.13$\times$ than DeepPrefetcher. Overall, SGDPs show much higher efficiency and practicality for real deployment applications.

\section{Conclusions}
To improve the performance of the data prefetcher in practice, this paper proposed SGDP, a novel stream-graph-based data prefetcher. SGDP takes each LBA delta stream as a weighted directed graph fusing both sequential and global features. By gated GNN and attention mechanism, SGDP extracts and aggregates the sequential and global information for better data prefetching. The experiment results from eight different real-world datasets demonstrate that SGDP outperforms SOTA methods and high speeds up inference time. The generalized SGDP variants can further adapt to extensive application scenarios. This novel data prefetcher has been verified in commercial hybrid storage systems in the experimental phase and will be deployed in the future product series.

\bibliographystyle{IEEEtran}
\bibliography{sample-base.bib}

\clearpage


\section*{Appendix A:  All Results of  Rolling Prediction Based Multi-Step Prefetching} 
\begin{figure*}[t]
\includegraphics[width=1.0\textwidth]{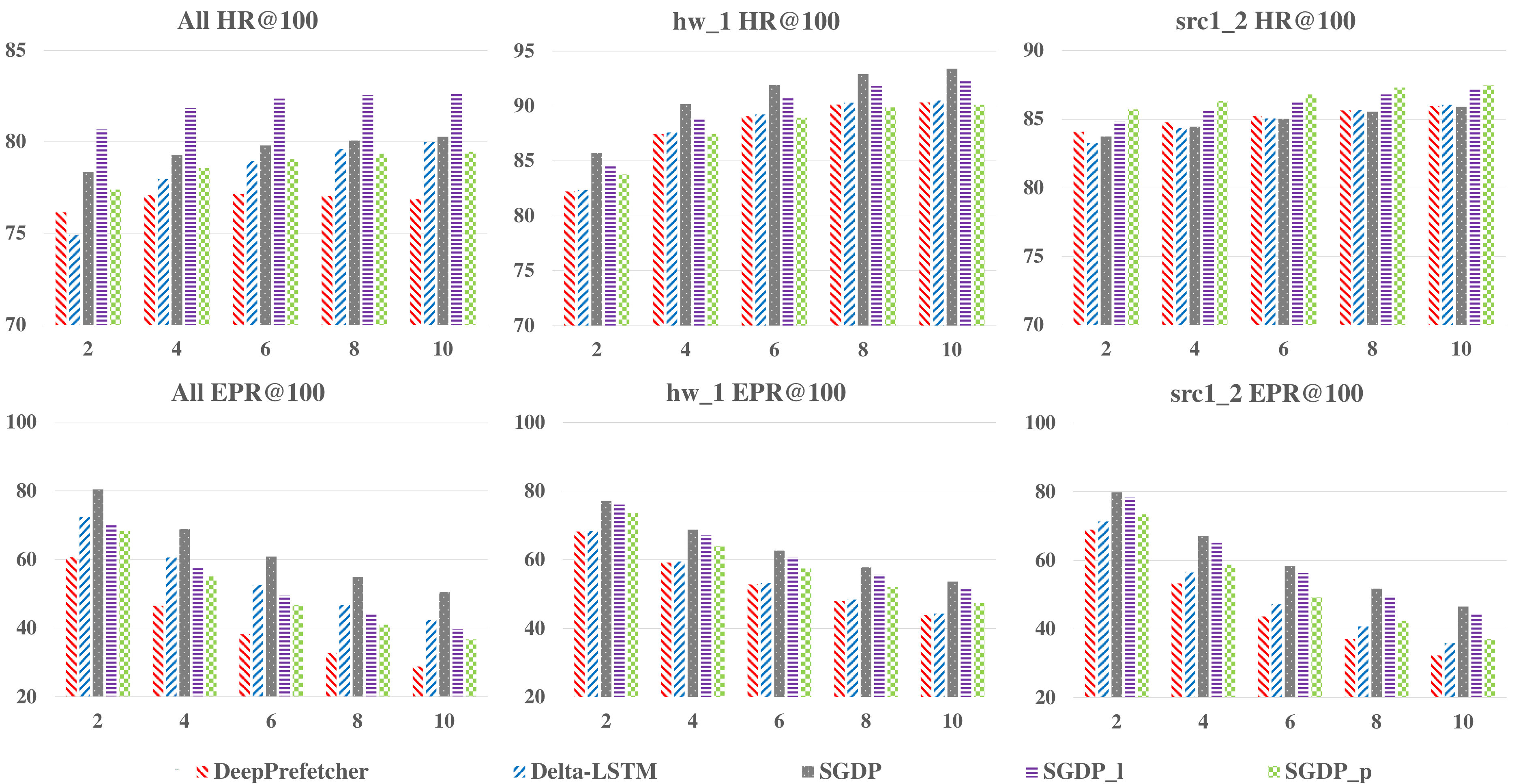}
\caption{Visualized Results of Multi-step Prefetching }
\label{Fig.all} 
\end{figure*}

We summarize  the  detailed  results of multi-step prefetching  in \textbf{Table} \ref{tab:multi}. As reported in the \textbf{Table} \ref{tab:multi}, SGDP and its variants achieve the best results in 80 cases    and 52  cases in terms of HR@100 and EPR@100, respectively. We also visualize the  results of the averaged results of all datasets and  two representative datasets as shown in Fig. \ref{Fig.all}. On hw\_1 dataset, SGDP stably performs the best with both the highest HR@100 and EPR@100. On src1\_2, SGDP$_{p}$ maintains highest HR@100 while SGDP have highest EPR@100.

\section*{Appendix B: Inference Efficiency}
To verify the practicality of SGDP compared to the SOTA methods, we collect statistics of inference time and report the detailed results of the number of LBA delta predictions that can be inferred per second by learning-based methods  in \textbf{Table} \ref{tab:efficiency}.   
SGDP and its variants process 469 to 670 LBA deltas  per second, while Delta-LSTM and DeepPrefetcher are only able to process 91.2 and 197.7 on average respectively. SGDP$_{l}$ speed up inference time up to 3.13 times than DeepPrefetcher. Overall, SGDP and its variants show much higher efficiency and practicality.

\section*{Appendix C: All Results of Single-step Prefetching Based on Different Cache Sizes}
We summarize  the  results of single-step prefetching about all eight datasets based on 20 different cache sizes ($\{5,10,20, \cdots ,90,100,200, \cdots ,900,1000\}$) from \textbf{Table} \ref{tab:hw_1} to \textbf{Table}\ref{tab:src1_2}. As reported in the Tables, SGDP and its variants (SGDP$_l$ and SGDP$_p$) achieve the best results in all 160 cases and 83 cases in terms of HR and EPR, respectively. As for EPR, our models are not far from the maximum (mostly within 1\%) in the non-first case.

\section*{Appendix D: Results and Algorithm Reproducing}
In data prefetch, researchers rarely open-source their code. We have contacted most authors for their baseline codes and benchmarks, but the response is almost non-existent, which makes it difficult for us to compare baselines. We could not find the related source code for graph-based methods. However, we can guarantee that the two learning-based methods we compared are the best methods available. They achieve better results than all graph-based ones using the same datasets, so we choose them as the baselines. Although the authors of the two methods did not give us the code directly, we received confirmation and positive feedback from them on our reproduction. So, we are confident that our results are now SOTA and definitely better than all the previous graph-based methods.

As for the discussed time-series-based methods in our paper, i.e., ARIMA and Informer, they are often discussed as baselines described in our Related Work section, it is reasonable to use ARIMA and Informer as baselines. As the other researchers said in their paper, those time-series-based methods perform well in datasets which have more sequence access patterns.

Researchers in this field hardly fully open-source their code, which not only makes it difficult for us to reproduce their methods but also hinders the development of the field. Therefore, we sincerely hope to promote the openness and development of the storage field and help more developers and researchers enter the community more efficiently by making our source code, datasets, and our reproduced and validated baselines available\footnote{Our codes are available at https://github.com/yyysjz1997/SGDP/.}.

\begin{table*}[!ht]
\centering
\caption{\textbf{Results of Multi-step Prefetching.} The results are in percentage, and the best results are highlighted in \textbf{bold}.}
\renewcommand{\arraystretch}{0.9}
\small 
\resizebox{1.0\textwidth}{!}{
\begin{tabular}{c|r|cccccccccc|cccccccccc}
\hline
\hline
      & Metrics & \multicolumn{10}{c|}{HR@100}                                                  & \multicolumn{10}{c}{EPR@100} \bigstrut\\
\hline
Dataset & \diagbox{Methods}{Steps} & 1     & 2     & 3     & 4     & 5     & 6     & 7     & 8     & 9     & 10    & 1     & 2     & 3     & 4     & 5     & 6     & 7     & 8     & 9     & 10 \bigstrut\\
\hline
\multirow{5}[4]{*}{hw\_1} & DeepPrefetcher & 74.6  & 82.2  & 85.3  & 87.4  & 88.4  & 89.1  & 89.7  & 90.1  & 90.2  & 90.3  & 75.9  & 68.2  & 63.0  & 59.2  & 55.8  & 52.8  & 50.3  & 48.0  & 45.8  & 43.9  \bigstrut[t]\\
      & Delta-LSTM & 74.8  & 82.3  & 85.4  & 87.6  & 88.6  & 89.2  & 89.9  & 90.3  & 90.4  & 90.5  & 76.0  & 68.4  & 63.2  & 59.5  & 56.1  & 53.2  & 50.6  & 48.4  & 46.2  & 44.3  \bigstrut[b]\\
\cline{2-22}      & SGDP  & \textbf{79.5} & \textbf{85.7} & \textbf{88.3} & \textbf{90.2} & \textbf{91.2} & \textbf{91.9} & \textbf{92.5} & \textbf{92.9} & \textbf{93.2} & \textbf{93.4} & \textbf{83.5} & \textbf{77.1} & \textbf{72.4} & \textbf{68.8} & \textbf{65.5} & \textbf{62.6} & \textbf{60.1} & \textbf{57.8} & \textbf{55.6} & \textbf{53.6} \bigstrut[t]\\
      & SGDP$_{l}$ & 78.8  & 84.7  & 87.1  & 88.8  & 89.9  & 90.8  & 91.3  & 91.8  & 92.1  & 92.3  & 82.7  & 76.1  & 71.0  & 67.1  & 63.7  & 60.8  & 58.1  & 55.8  & 53.6  & 51.5  \\
      & SGDP$_{p}$ & 78.2  & 83.8  & 86.0  & 87.4  & 88.3  & 88.9  & 89.5  & 89.9  & 90.1  & 90.1  & 80.4  & 73.6  & 68.1  & 64.0  & 60.5  & 57.4  & 54.7  & 52.1  & 49.6  & 47.3  \bigstrut[b]\\
\hline
\multirow{5}[4]{*}{hw\_2} & DeepPrefetcher & 92.5  & 93.4  & 93.7  & 93.9  & 94.0  & 94.1  & 94.1  & 94.1  & 94.2  & 94.2  & 94.0  & 89.6  & 85.7  & 82.2  & 79.0  & 76.0  & 73.3  & 70.7  & 68.3  & 66.0  \bigstrut[t]\\
      & Delta-LSTM & 92.8  & 93.6  & 94.0  & 94.3  & 94.4  & 94.5  & 94.5  & 94.6  & 94.6  & 94.6  & 94.2  & 89.9  & 86.2  & 83.0  & 80.0  & 77.2  & 74.7  & 72.4  & 70.2  & 68.1  \bigstrut[b]\\
\cline{2-22}      & SGDP  & 93.0  & 94.1  & 94.5  & 94.8  & 94.9  & 95.0  & 95.1  & 95.1  & 95.2  & 95.2  & \textbf{97.7} & \textbf{96.0} & \textbf{94.7} & \textbf{93.4} & \textbf{92.3} & \textbf{91.3} & \textbf{90.4} & \textbf{89.4} & \textbf{88.5} & \textbf{87.7} \bigstrut[t]\\
      & SGDP$_{l}$ & 93.1  & 94.1  & 94.6  & 94.8  & 95.0  & 95.1  & 95.2  & 95.2  & 95.3  & 95.3  & 97.2  & 95.2  & 93.3  & 91.7  & 90.1  & 88.7  & 87.3  & 85.9  & 84.6  & 83.4  \\
      & SGDP$_{p}$ & \textbf{94.0} & \textbf{95.1} & \textbf{95.5} & \textbf{95.7} & \textbf{95.9} & \textbf{96.0} & \textbf{96.0} & \textbf{96.0} & \textbf{96.1} & \textbf{96.1} & 95.0  & 91.2  & 88.0  & 85.0  & 82.3  & 79.7  & 77.4  & 75.0  & 72.9  & 70.8  \bigstrut[b]\\
\hline
\multirow{5}[4]{*}{hw\_3} & DeepPrefetcher & 50.7  & 51.8  & 51.8  & 51.8  & 51.8  & 51.8  & 51.8  & 51.8  & 51.8  & 51.8  & 50.7  & 34.9  & 26.3  & 21.1  & 17.7  & 15.2  & 13.3  & 11.8  & 10.6  & 9.7  \bigstrut[t]\\
      & Delta-LSTM & 56.8  & 60.7  & 67.8  & 68.3  & 68.4  & 68.5  & 68.4  & 68.4  & 68.4  & 68.4  & 66.8  & 52.4  & 46.9  & 41.1  & 36.5  & 32.8  & 29.7  & 27.1  & 25.3  & 23.5  \bigstrut[b]\\
\cline{2-22}      & SGDP  & 76.6  & 77.7  & 78.0  & 78.0  & 78.0  & 78.0  & 78.0  & 78.0  & 78.1  & 78.0  & \textbf{89.5} & \textbf{81.6} & \textbf{74.4} & \textbf{68.9} & \textbf{63.8} & \textbf{59.9} & \textbf{56.1} & \textbf{53.1} & \textbf{50.8} & \textbf{48.2} \bigstrut[t]\\
      & SGDP$_{l}$ & \textbf{79.0} & \textbf{80.2} & \textbf{80.5} & \textbf{80.6} & \textbf{80.8} & \textbf{80.8} & \textbf{80.8} & \textbf{80.8} & \textbf{80.9} & \textbf{80.8} & 84.2  & 73.8  & 65.3  & 58.9  & 53.5  & 49.2  & 45.3  & 42.2  & 39.5  & 37.0  \\
      & SGDP$_{p}$ & 48.3  & 48.5  & 48.5  & 48.6  & 48.6  & 48.7  & 48.7  & 48.9  & 49.2  & 49.3  & 73.1  & 61.6  & 53.4  & 47.5  & 42.9  & 39.2  & 36.2  & 34.1  & 32.8  & 31.5  \bigstrut[b]\\
\hline
\multirow{5}[4]{*}{hm\_1} & DeepPrefetcher & 59.1  & 60.0  & 59.2  & 58.2  & 57.1  & 56.1  & 55.2  & 54.4  & 53.6  & 53.0  & 56.0  & 39.6  & 29.8  & 23.5  & 19.1  & 16.0  & 13.7  & 11.9  & 10.5  & 9.4  \bigstrut[t]\\
      & Delta-LSTM & 50.6  & 55.5  & 56.7  & 57.8  & 58.7  & 59.3  & 59.9  & 60.2  & 60.4  & 60.6  & 72.8  & 63.5  & 54.9  & 48.6  & 43.6  & 39.6  & 36.3  & 33.5  & 31.5  & 29.5  \bigstrut[b]\\
\cline{2-22}      & SGDP  & 55.7  & 55.8  & 55.5  & 55.3  & 55.1  & 54.9  & 54.6  & 54.3  & 54.1  & 54.0  & \textbf{90.1} & \textbf{80.5} & \textbf{72.0} & \textbf{64.9} & \textbf{59.2} & \textbf{54.2} & \textbf{50.0} & \textbf{46.6} & \textbf{44.8} & \textbf{42.3} \bigstrut[t]\\
      & SGDP$_{l}$ & 61.4  & 63.1  & 63.6  & 63.7  & 63.4  & 63.1  & 62.8  & 62.4  & 62.1  & 61.8  & 60.8  & 45.1  & 35.8  & 29.7  & 25.2  & 21.9  & 19.3  & 17.2  & 15.5  & 14.1  \\
      & SGDP$_{p}$ & \textbf{62.9} & \textbf{65.0} & \textbf{65.7} & \textbf{65.8} & \textbf{65.8} & \textbf{65.4} & \textbf{65.1} & \textbf{64.8} & \textbf{64.5} & \textbf{64.1} & 63.8  & 48.6  & 39.2  & 32.7  & 28.0  & 24.3  & 21.5  & 19.2  & 17.4  & 15.8  \bigstrut[b]\\
\hline
\multirow{5}[4]{*}{mds\_0} & DeepPrefetcher & 73.7  & 81.3  & 81.7  & 81.8  & 81.6  & 81.2  & 80.9  & 80.6  & 80.3  & 79.9  & 77.5  & 61.3  & 50.5  & 42.6  & 36.3  & 31.2  & 27.4  & 24.4  & 21.9  & 19.9  \bigstrut[t]\\
      & Delta-LSTM & 69.6  & 75.6  & 77.2  & 78.2  & 79.0  & 79.7  & 80.3  & 80.9  & 81.4  & 81.8  & \textbf{87.8} & \textbf{80.7} & \textbf{74.6} & \textbf{69.4} & \textbf{65.2} & \textbf{61.4} & \textbf{58.1} & \textbf{55.2} & \textbf{52.7} & \textbf{50.4} \bigstrut[b]\\
\cline{2-22}      & SGDP  & 76.3  & 76.9  & 77.4  & 77.8  & 78.1  & 78.5  & 78.9  & 79.1  & 79.4  & 79.5  & 87.0  & 78.1  & 71.1  & 65.4  & 60.7  & 56.6  & 53.2  & 50.3  & 47.7  & 45.3  \bigstrut[t]\\
      & SGDP$_{l}$ & 77.5  & 78.9  & 79.5  & 79.8  & 80.0  & 80.1  & 80.2  & 80.3  & 80.3  & 80.3  & 73.9  & 59.7  & 49.8  & 43.0  & 37.8  & 33.6  & 30.4  & 27.8  & 25.5  & 23.5  \\
      & SGDP$_{p}$ & \textbf{79.8} & \textbf{82.0} & \textbf{82.9} & \textbf{83.6} & \textbf{84.0} & \textbf{84.3} & \textbf{84.6} & \textbf{84.8} & \textbf{85.0} & \textbf{85.0} & 81.9  & 71.3  & 63.2  & 56.9  & 51.5  & 47.1  & 43.3  & 40.2  & 37.3  & 34.7  \bigstrut[b]\\
\hline
\multirow{5}[4]{*}{proj\_0} & DeepPrefetcher & 79.1  & 81.3  & 81.7  & 82.0  & 82.2  & 82.2  & 82.3  & 82.2  & 82.3  & 82.2  & 78.6  & 64.6  & 55.1  & 47.9  & 42.4  & 37.8  & 34.2  & 31.1  & 28.5  & 26.4  \bigstrut[t]\\
      & Delta-LSTM & 69.1  & 76.9  & 78.4  & 79.1  & 79.5  & 79.9  & 80.2  & 80.5  & 80.7  & 80.9  & 86.2  & \textbf{79.6} & \textbf{72.7} & \textbf{67.1} & \textbf{62.3} & \textbf{58.2} & \textbf{54.6} & \textbf{51.5} & \textbf{48.8} & \textbf{46.3} \bigstrut[b]\\
\cline{2-22}      & SGDP  & 78.5  & 78.9  & 79.1  & 79.3  & 79.5  & 79.7  & 79.8  & 79.9  & 80.1  & 80.1  & \textbf{87.6} & 78.2  & 70.7  & 64.5  & 59.5  & 55.1  & 51.4  & 48.1  & 45.4  & 42.9  \bigstrut[t]\\
      & SGDP$_{l}$ & 81.1  & 81.9  & 82.3  & 82.6  & 82.9  & \textbf{83.1} & \textbf{83.2} & 83.3  & \textbf{83.5} & \textbf{83.5} & 83.6  & 72.4  & 64.0  & 57.4  & 52.1  & 47.8  & 44.1  & 41.0  & 38.3  & 36.0  \\
      & SGDP$_{p}$ & \textbf{81.3} & \textbf{82.1} & \textbf{82.4} & \textbf{82.7} & \textbf{82.9} & 83.1  & 83.1  & \textbf{83.3} & 83.5  & 83.5  & 80.2  & 67.7  & 58.7  & 51.9  & 46.5  & 42.1  & 38.5  & 35.5  & 32.9  & 30.6  \bigstrut[b]\\
\hline
\multirow{5}[4]{*}{prxy\_0} & DeepPrefetcher & 70.2  & 75.0  & 76.2  & 76.8  & 77.2  & 77.5  & 77.6  & 77.5  & 77.6  & 77.6  & 70.4  & 58.4  & 49.3  & 42.6  & 37.5  & 33.5  & 30.1  & 27.3  & 25.0  & 23.1  \bigstrut[t]\\
      & Delta-LSTM & 64.2  & 71.5  & 73.2  & 74.1  & 74.8  & 75.4  & 76.1  & 76.4  & 76.7  & 77.0  & 79.3  & \textbf{72.8} & \textbf{65.6} & \textbf{59.9} & \textbf{55.4} & \textbf{51.4} & \textbf{48.3} & \textbf{45.3} & \textbf{42.9} & \textbf{40.7} \bigstrut[b]\\
\cline{2-22}      & SGDP  & 73.2  & 73.9  & 74.4  & 74.6  & 75.3  & 75.5  & 75.6  & 75.7  & 76.1  & 76.1  & \textbf{83.3} & 72.3  & 64.3  & 58.0  & 53.2  & 49.0  & 45.5  & 42.5  & 40.1  & 37.6  \bigstrut[t]\\
      & SGDP$_{l}$ & \textbf{76.5} & \textbf{77.7} & \textbf{78.3} & \textbf{78.7} & \textbf{79.4} & \textbf{79.6} & \textbf{79.8} & \textbf{79.9} & \textbf{80.2} & \textbf{80.2} & 74.3  & 61.4  & 53.1  & 47.0  & 42.5  & 38.6  & 35.5  & 32.8  & 30.7  & 28.6  \\
      & SGDP$_{p}$ & 76.2  & 77.2  & 77.8  & 78.2  & 79.0  & 79.3  & 79.6  & 79.8  & 80.1  & 80.1  & 72.6  & 59.1  & 50.4  & 44.0  & 39.5  & 35.8  & 32.7  & 30.1  & 28.0  & 26.0  \bigstrut[b]\\
\hline
\multirow{5}[4]{*}{src1\_2} & DeepPrefetcher & 82.9  & 84.1  & 84.5  & 84.8  & 85.0  & 85.2  & 85.5  & 85.6  & 85.9  & 86.0  & 80.9  & 68.9  & 60.0  & 53.3  & 48.0  & 43.7  & 40.1  & 37.1  & 34.5  & 32.3  \bigstrut[t]\\
      & Delta-LSTM & 79.6  & 83.3  & 84.0  & 84.4  & 84.7  & 85.1  & 85.4  & 85.7  & 85.9  & 86.0  & 81.3  & 71.3  & 63.0  & 56.4  & 51.3  & 47.2  & 43.7  & 40.7  & 38.1  & 35.9  \bigstrut[b]\\
\cline{2-22}      & SGDP  & 83.1  & 83.7  & 84.1  & 84.5  & 84.8  & 85.0  & 85.4  & 85.5  & 85.8  & 85.9  & \textbf{88.5} & \textbf{79.8} & \textbf{72.8} & \textbf{67.1} & \textbf{62.4} & \textbf{58.3} & \textbf{54.9} & \textbf{51.7} & \textbf{49.0} & \textbf{46.5} \bigstrut[t]\\
      & SGDP$_{l}$ & 83.9  & 84.8  & 85.3  & 85.7  & 86.0  & 86.3  & 86.7  & 86.8  & 87.1  & 87.2  & 87.4  & 78.3  & 71.0  & 65.1  & 60.4  & 56.3  & 52.8  & 49.7  & 47.0  & 44.5  \\
      & SGDP$_{p}$ & \textbf{84.8} & \textbf{85.7} & \textbf{86.0} & \textbf{86.3} & \textbf{86.6} & \textbf{86.8} & \textbf{87.0} & \textbf{87.3} & \textbf{87.5} & \textbf{87.5} & 84.0  & 73.4  & 65.2  & 58.7  & 53.6  & 49.2  & 45.5  & 42.3  & 39.5  & 37.0  \bigstrut[b]\\
\hline
\hline
\end{tabular}%

}
\label{tab:multi}%
\end{table*}%

\begin{table*}[!h]
	\centering
	\footnotesize
	\caption{The number of LBA delta predictions that can be inferred per second by learning-based methods.}
	\renewcommand{\arraystretch}{0.6}
	\resizebox{\textwidth}{!}{
        \begin{tabular}{r|rrrrrrrr|c}
        \hline \hline
        \diagbox{Method}{dataset} & hw\_1  & hw\_2  & hw\_3  & hm\_1  & mds\_0 & proj\_0 & prxy\_0 & src1\_2 & avg    \\ \hline 
        Delta-LSTM & 89.4 & 87.4 & 94.5 & 92.4 & 90.7 & 91.5 & 88.4 & 95.1 & 91.2 \\
        DeepPrefetcher & 208.2 & 154.5 & 194.2 & 160.1 & 248.4 & 178.4 & 187.9 & 249.6 & 197.7 \\ \hline 
        SGDP & 644.5 & 692.4 & 666.1 & 515.2 & 543.5 & 553.3 & 470.0 & 550.7 & 579.5 \\
        SGDP$_{l}$ & 634.7 & 686.9 & 614.7 & 500.1 & 651.4 & 663.9 & 526.3 & 670.7 & 618.6 \\
        SGDP$_{p}$ & 599.5 & 645.6 & 593.9 & 567.0 & 491.7 & 529.3 & 574.8 & 558.7 & 570.1 \\ \hline \hline
        \end{tabular}
        }
	\label{tab:efficiency}
\end{table*}


\begin{table*}[!h]
	\centering
	\caption{Results of single-step prefetching based on different cache sizes about dataset \textbf{hw\_1}.}
	\renewcommand{\arraystretch}{1.2}
	\resizebox{\textwidth}{!}{
    \begin{tabular}{rcccccccccccccccccccc}
    \hline \hline
    \multicolumn{21}{c}{HR@N} \\ \hline
    \multicolumn{1}{c|}{\diagbox{Methods}{Cache sizes}} & 5 & 10 & 20 & 30 & 40 & 50 & 60 & 70 & 80 & 90 & 100 & 200 & 300 & 400 & 500 & 600 & 700 & 800 & 900 & 1000 \\ \hline
    \multicolumn{1}{r|}{No\_pre} & 0 & 0 & 0 & 0 & 0 & 0 & 0 & 0.3 & 0.3 & 0.3 & 0.3 & 0.4 & 1.7 & 2.6 & 2.7 & 3.1 & 7.6 & 15.6 & 38.9 & 54.2 \\
    \multicolumn{1}{r|}{Naive} & 56.9 & 57.5 & 57.8 & 57.9 & 57.9 & 57.9 & 57.9 & 57.9 & 58.0 & 58.0 & 58.0 & 58.1 & 58.2 & 58.9 & 59.2 & 59.5 & 59.7 & 59.9 & 61.6 & 63.2 \\
    \multicolumn{1}{r|}{Stride} & 43.6 & 43.7 & 43.8 & 43.9 & 43.9 & 43.9 & 43.9 & 44.0 & 44.0 & 44.0 & 44.0 & 44.0 & 44.6 & 45.4 & 45.7 & 46.0 & 46.2 & 50.2 & 53.8 & 65.8 \\
    \multicolumn{1}{r|}{ARIMA} & 1.5 & 1.9 & 2.3 & 2.6 & 3.0 & 3.1 & 3.1 & 3.4 & 3.8 & 3.9 & 4.0 & 4.3 & 4.5 & 5.0 & 6.0 & 6.7 & 7.6 & 8.0 & 8.3 & 8.8 \\
    \multicolumn{1}{r|}{Informer} & 0.2 & 0.2 & 0.4 & 0.5 & 0.5 & 0.6 & 0.7 & 0.7 & 0.8 & 0.8 & 0.9 & 1.3 & 1.6 & 2.0 & 2.4 & 3.9 & 4.4 & 5.2 & 5.5 & 5.8 \\
    \multicolumn{1}{r|}{DeepPrefetcher} & 73.9  & 74.3  & 74.5  & 74.6  & 74.6  & 74.6  & 74.6  & 74.6  & 74.6  & 74.6  & 74.6  & 74.7  & 74.8  & 75.3  & 75.7  & 75.7  & 75.9  & 76.5  & 77.8  & 79.2 \\
    \multicolumn{1}{r|}{Delta-LSTM} & 74.0 & 74.4 & 74.6 & 74.7 & 74.7 & 74.7 & 74.7 & 74.7 & 74.8 & 74.8 & 74.8 & 74.8 & 74.9 & 75.4 & 75.8 & 75.8 & 76.0 & 76.6 & 77.9 & 79.3 \\ \hline
    \multicolumn{1}{r|}{SGDP} & \textbf{78.7}& \textbf{79.2} & \textbf{79.4} & \textbf{79.4} & \textbf{79.4} & \textbf{79.5}& \textbf{79.5}& \textbf{79.5} & \textbf{79.5} & \textbf{79.5} & \textbf{79.5} & \textbf{79.5} & \textbf{79.6} & \textbf{79.9} & \textbf{80.2} & \textbf{80.3} & \textbf{80.4} & \textbf{81.3} & \textbf{82.4} & \textbf{85.8}\\
    \multicolumn{1}{r|}{SGDP$_l$} & 77.9 & 78.5 & 78.7 & 78.8 & 78.8 & 78.8 & 78.8 & 78.8 & 78.8 & 78.8 & 78.8 & 78.9 & 79.0 & 79.3 & 79.6 & 79.6 & 79.8 & 80.7 & 81.7 & 84.9 \\
    \multicolumn{1}{r|}{SGDP$_p$} & 74.9 & 75.7 & 76.7 & 77.1 & 77.4 & 77.6 & 77.7 & 77.9 & 78.0 & 78.1 & 78.2 & 78.5 & 78.6 & 79.0 & 79.3 & 79.4 & 79.5 & 80.2 & 81.1 & 83.6 \\ \hline
    \multicolumn{21}{c}{EPR@N} \\ \hline
    \multicolumn{1}{c|}{\diagbox{Methods}{Cache sizes}}& 5 & 10 & 20 & 30 & 40 & 50 & 60 & 70 & 80 & 90 & 100 & 200 & 300 & 400 & 500 & 600 & 700 & 800 & 900 & 1000 \\ \hline
    \multicolumn{1}{r|}{Naive} & 61.1 & 63.3 & 64.3 & 64.4 & 64.4 & 64.4 & 64.5 & 64.5 & 64.5 & 64.5 & 64.5 & 64.6 & 64.8 & 64.9 & 65.0 & 65.1 & 65.1 & 65.2 & 64.7 & 64.5 \\
    \multicolumn{1}{r|}{Stride} & 80.2 & 80.5 & 81.0 & 81.0 & 81.0 & 81.1 & 81.1 & 81.1 & 81.1 & 81.1 & 81.1 & 81.1 & 81.2 & 81.3 & 81.4 & 81.4 & 81.5 & 81.3 & 81.1 & 80.6 \\
    \multicolumn{1}{r|}{ARIMA} & 1.5 & 1.9 & 2.5 & 2.8 & 3.2 & 3.3 & 3.3 & 3.6 & 4.1 & 4.2 & 4.3 & 4.4 & 4.7 & 5.1 & 5.3 & 5.6 & 5.7 & 6.0 & 6.2 & 6.2 \\
    \multicolumn{1}{r|}{Informer} & 0.2 & 0.3 & 0.4 & 0.5 & 0.6 & 0.6 & 0.7 & 0.7 & 0.8 & 0.8 & 0.9 & 1.1 & 1.4 & 1.7 & 2.1 & 2.3 & 2.6 & 2.8 & 2.8 & 2.9 \\
    \multicolumn{1}{r|}{DeepPrefetcher} & 74.6  & 75.4  & 75.8  & 75.8  & 75.8  & 75.8  & 75.9  & 75.9  & 75.9  & 75.9  & 75.9  & 75.9  & 76.0  & 76.2  & 76.4  & 76.5  & 76.5  & 76.5  & 76.5  & 76.5 \\
    \multicolumn{1}{r|}{Delta-LSTM} & 74.7 & 75.5 & 75.9 & 75.9 & 75.9 & 76.0 & 76.0 & 76.0 & 76.0 & 76.0 & 76.0 & 76.0 & 76.1 & 76.3 & 76.5 & 76.6 & 76.6 & 76.6 & 76.6 & 76.6 \\ \hline
    \multicolumn{1}{r|}{SGDP} & \textbf{81.9} & \textbf{82.9} & \textbf{83.4} &\textbf{ 83.4} & \textbf{83.4} & \textbf{83.5} & \textbf{83.5} & \textbf{83.5} & \textbf{83.5} & \textbf{83.5} & \textbf{83.5} & \textbf{83.5} & \textbf{83.6} & \textbf{83.6}&\textbf{ 83.7} & \textbf{83.7} & \textbf{83.7} & \textbf{83.4} & \textbf{83.0}& \textbf{81.6} \\
    \multicolumn{1}{r|}{SGDP$_l$} & 80.9 & 82.1 & 82.6 & 82.6 & 82.6 & 82.7 & 82.7 & 82.7 & 82.7 & 82.7 & 82.7 & 82.8 & 82.9 & 82.9 & 83.0 & 83.0 & 83.1 & 82.7 & 82.3 & 80.6 \\
    \multicolumn{1}{r|}{SGDP$_p$} & 76.4 & 77.6 & 78.8 & 79.2 & 79.6 & 79.8 & 80.0 & 80.1 & 80.2 & 80.3 & 80.4 & 80.8 & 80.9 & 81.0 & 81.1 & 81.2 & 81.2 & 80.9 & 80.7 & 79.6 \\ \hline \hline
    \end{tabular}   
        }
	\label{tab:hw_1}
\end{table*}

\begin{table*}[!h]
	\centering
	\caption{Results of single-step prefetching based on different cache sizes about dataset \textbf{hw\_2}.}
	\renewcommand{\arraystretch}{1.2}
	\resizebox{\textwidth}{!}{
    \begin{tabular}{rllllllllllllllllllll}
\hline \hline
\multicolumn{21}{c}{HR@N} \\ \hline
    \multicolumn{1}{c|}{\diagbox{Methods}{Cache sizes}} & \multicolumn{1}{c}{5} & \multicolumn{1}{c}{10} & \multicolumn{1}{c}{20} & \multicolumn{1}{c}{30} & \multicolumn{1}{c}{40} & \multicolumn{1}{c}{50} & \multicolumn{1}{c}{60} & \multicolumn{1}{c}{70} & \multicolumn{1}{c}{80} & \multicolumn{1}{c}{90} & \multicolumn{1}{c}{100} & \multicolumn{1}{c}{200} & \multicolumn{1}{c}{300} & \multicolumn{1}{c}{400} & \multicolumn{1}{c}{500} & \multicolumn{1}{c}{600} & \multicolumn{1}{c}{700} & \multicolumn{1}{c}{800} & \multicolumn{1}{c}{900} & \multicolumn{1}{c}{1000} \\ \hline
\multicolumn{1}{r|}{No\_pre} & 1.0 & 1.0 & 1.0 & 1.1 & 1.1 & 1.1 & 1.1 & 1.1 & 1.1 & 1.1 & 1.1 & 1.1 & 1.1 & 1.1 & 1.1 & 1.1 & 1.1 & 1.1 & 1.1 & 1.1 \\
\multicolumn{1}{r|}{Naive} & 92.3 & 92.5 & 92.5 & 92.5 & 92.6 & 92.6 & 92.6 & 92.6 & 92.6 & 92.6 & 92.6 & 92.7 & 92.7 & 92.7 & 92.7 & 92.7 & 92.7 & 92.7 & 92.7 & 92.7 \\
\multicolumn{1}{r|}{Stride} & 91.0 & 91.0 & 91.0 & 91.0 & 91.1 & 91.1 & 91.1 & 91.1 & 91.1 & 91.1 & 91.1 & 91.1 & 91.1 & 91.1 & 91.1 & 91.1 & 91.1 & 91.1 & 91.1 & 91.1 \\
\multicolumn{1}{r|}{ARIMA} & 82.6 & 82.8 & 82.8 & 82.8 & 82.8 & 82.9 & 82.9 & 82.9 & 82.9 & 82.9 & 82.9 & 82.9 & 83.0 & 83.0 & 83.0 & 83.0 & 83.0 & 83.0 & 83.0 & 83.0 \\
\multicolumn{1}{r|}{Informer} & 1.0 & 1.0 & 1.0 & 1.0 & 1.0 & 1.0 & 1.1 & 1.1 & 1.1 & 1.1 & 1.1 & 1.1 & 1.1 & 1.1 & 1.1 & 1.1 & 1.1 & 1.1 & 1.1 & 1.1 \\
\multicolumn{1}{r|}{DeepPrefetcher}& 92.2  & 92.2  & 92.3  & 92.4  & 92.4  & 92.5  & 92.5  & 92.5  & 92.5  & 92.5  & 92.5  & 92.6  & 92.7  & 92.7  & 92.7  & 92.7  & 92.8  & 92.8  & 92.8  & 92.8 \\
\multicolumn{1}{r|}{Delta-LSTM} & 92.4 & 92.5 & 92.6 & 92.6 & 92.7 & 92.7 & 92.7 & 92.8 & 92.8 & 92.8 & 92.8 & 92.9 & 92.9 & 92.9 & 92.9 & 93.0 & 93.0 & 93.0 & 93.0 & 93.1 \\ \hline
\multicolumn{1}{r|}{SGDP} & 92.9 & 93.0 & 93.0 & 93.0 & 93.0 & 93.0 & 93.0 & 93.0 & 93.0 & 93.0 & 93.0 & 93.1 & 93.1 & 93.1 & 93.1 & 93.1 & 93.1 & 93.1 & 93.1 & 93.1 \\
\multicolumn{1}{r|}{SGDP$_l$} & 92.9 & 92.9 & 93.0 & 93.0 & 93.0 & 93.0 & 93.0 & 93.1 & 93.1 & 93.1 & 93.1 & 93.1 & 93.1 & 93.1 & 93.1 & 93.1 & 93.2 & 93.2 & 93.2 & 93.2 \\
\multicolumn{1}{r|}{SGDP$_p$} & \textbf{93.6} & \textbf{93.7} &\textbf{93.8} & \textbf{93.9} & \textbf{94.0} &\textbf{ 94.0} & \textbf{94.0} &\textbf{94.0}& \textbf{94.0} &\textbf{94.0} & \textbf{94.0} & \textbf{94.1}& \textbf{94.1} & \textbf{94.1} & \textbf{94.1} & \textbf{94.2} & \textbf{94.2} & \textbf{94.2} & \textbf{94.2} & \textbf{94.2} \\ \hline
\multicolumn{21}{c}{EPR@N} \\ \hline
    \multicolumn{1}{c|}{\diagbox{Methods}{Cache sizes}} & \multicolumn{1}{c}{5} & \multicolumn{1}{c}{10} & \multicolumn{1}{c}{20} & \multicolumn{1}{c}{30} & \multicolumn{1}{c}{40} & \multicolumn{1}{c}{50} & \multicolumn{1}{c}{60} & \multicolumn{1}{c}{70} & \multicolumn{1}{c}{80} & \multicolumn{1}{c}{90} & \multicolumn{1}{c}{100} & \multicolumn{1}{c}{200} & \multicolumn{1}{c}{300} & \multicolumn{1}{c}{400} & \multicolumn{1}{c}{500} & \multicolumn{1}{c}{600} & \multicolumn{1}{c}{700} & \multicolumn{1}{c}{800} & \multicolumn{1}{c}{900} & \multicolumn{1}{c}{1000} \\ \hline
\multicolumn{1}{r|}{Naive} & 93.0 & 93.3 & 93.4 & 93.5 & 93.5 & 93.6 & 93.6 & 93.7 & 93.7 & 93.7 & 93.7 & 93.9 & 93.9 & 93.9 & 93.9 & 94.0 & 94.0 & 94.0 & 94.0 & 94.0 \\
\multicolumn{1}{r|}{Stride} &\textbf{99.1} & \textbf{99.1} &\textbf{99.1}& \textbf{99.1} &\textbf{99.2}& \textbf{99.2} & \textbf{99.2} & \textbf{99.2} & \textbf{99.2} & \textbf{99.2} & \textbf{99.2} & \textbf{99.2} & \textbf{99.2} & \textbf{99.2} & \textbf{99.2} & \textbf{99.2} & \textbf{99.2} & \textbf{99.2} & \textbf{99.2} & \textbf{99.2} \\
\multicolumn{1}{r|}{ARIMA} & 85.5 & 85.9 & 86.0 & 86.1 & 86.1 & 86.1 & 86.1 & 86.1 & 86.1 & 86.1 & 86.2 & 86.2 & 86.3 & 86.3 & 86.3 & 86.4 & 86.4 & 86.4 & 86.4 & 86.4 \\
\multicolumn{1}{r|}{Informer} & 0.0 & 0.0 & 0.0 & 0.0 & 0.0 & 0.0 & 0.0 & 0.0 & 0.0 & 0.0 & 0.0 & 0.0 & 0.0 & 0.0 & 0.0 & 0.0 & 0.0 & 0.0 & 0.0 & 0.0 \\
\multicolumn{1}{r|}{DeepPrefetcher} & 93.3  & 93.4  & 93.6  & 93.6  & 93.7  & 93.8  & 93.9  & 94.0  & 94.0  & 94.0  & 94.0  & 94.2  & 94.3  & 94.3  & 94.3  & 94.4  & 94.5  & 94.5  & 94.5  & 94.5 \\
\multicolumn{1}{r|}{Delta-LSTM} & 93.6 & 93.7 & 93.8 & 93.9 & 94.0 & 94.1 & 94.1 & 94.1 & 94.2 & 94.2 & 94.2 & 94.4 & 94.4 & 94.5 & 94.5 & 94.6 & 94.7 & 94.7 & 94.7 & 94.7 \\ \hline
\multicolumn{1}{r|}{SGDP} & 97.5 & 97.5 & 97.6 & 97.6 & 97.6 & 97.7 & 97.7 & 97.7 & 97.7 & 97.7 & 97.7 & 97.7 & 97.8 & 97.8 & 97.8 & 97.8 & 97.8 & 97.8 & 97.8 & 97.8 \\
\multicolumn{1}{r|}{SGDP$_l$} & 96.8 & 97.0 & 97.0 & 97.1 & 97.1 & 97.2 & 97.2 & 97.2 & 97.2 & 97.2 & 97.2 & 97.3 & 97.3 & 97.3 & 97.3 & 97.4 & 97.4 & 97.4 & 97.4 & 97.4 \\
\multicolumn{1}{r|}{SGDP$_p$} & 94.2 & 94.4 & 94.6 & 94.7 & 94.8 & 94.9 & 94.9 & 95.0 & 95.0 & 95.0 & 95.0 & 95.1 & 95.2 & 95.2 & 95.2 & 95.3 & 95.4 & 95.4 & 95.4 & 95.4 \\ \hline \hline
\end{tabular}
        }
	\label{tab:hw_2}
\end{table*}

\begin{table*}[!h]
	\centering
	\caption{Results of single-step prefetching based on different cache sizes about dataset \textbf{hw\_3}.}
	\renewcommand{\arraystretch}{1.3}
	\resizebox{\textwidth}{!}{
   \begin{tabular}{rllllllllllllllllllll}
\hline \hline 
\multicolumn{21}{c}{HR@N} \\ \hline
\multicolumn{1}{c|}{\diagbox{Methods}{Cache sizes}}  & \multicolumn{1}{c}{5} & \multicolumn{1}{c}{10} & \multicolumn{1}{c}{20} & \multicolumn{1}{c}{30} & \multicolumn{1}{c}{40} & \multicolumn{1}{c}{50} & \multicolumn{1}{c}{60} & \multicolumn{1}{c}{70} & \multicolumn{1}{c}{80} & \multicolumn{1}{c}{90} & \multicolumn{1}{c}{100} & \multicolumn{1}{c}{200} & \multicolumn{1}{c}{300} & \multicolumn{1}{c}{400} & \multicolumn{1}{c}{500} & \multicolumn{1}{c}{600} & \multicolumn{1}{c}{700} & \multicolumn{1}{c}{800} & \multicolumn{1}{c}{900} & \multicolumn{1}{c}{1000} \\ \hline
\multicolumn{1}{r|}{No\_pre} & 0.0 & 0.0 & 0.0 & 0.0 & 0.0 & 0.0 & 0.0 & 0.0 & 0.0 & 0.1 & 0.1 & 0.2 & 0.3 & 0.9 & 0.9 & 1.0 & 1.1 & 1.1 & 1.2 & 1.3 \\
\multicolumn{1}{r|}{Naive} & 47.6 & 47.7 & 47.8 & 47.8 & 47.8 & 47.8 & 47.8 & 47.8 & 47.9 & 47.9 & 47.9 & 48.2 & 48.2 & 48.2 & 48.2 & 48.2 & 48.7 & 48.8 & 48.8 & 48.8 \\
\multicolumn{1}{r|}{Stride} & 38.4 & 38.4 & 38.5 & 38.5 & 38.5 & 38.5 & 38.5 & 38.5 & 38.6 & 38.6 & 38.6 & 38.8 & 38.8 & 39.3 & 39.4 & 39.5 & 39.5 & 39.5 & 39.6 & 39.6 \\
\multicolumn{1}{r|}{ARIMA} & 0.1 & 0.3 & 0.3 & 0.3 & 0.3 & 0.3 & 0.3 & 0.3 & 0.3 & 0.3 & 0.3 & 0.4 & 0.5 & 0.5 & 0.5 & 0.6 & 0.6 & 1.2 & 1.2 & 1.3 \\
\multicolumn{1}{r|}{Informer} & 0.0 & 0.0 & 0.0 & 0.0 & 0.0 & 0.0 & 0.0 & 0.0 & 0.0 & 0.0 & 0.0 & 0.1 & 0.2 & 0.2 & 0.2 & 0.3 & 0.3 & 0.9 & 0.9 & 0.9 \\
\multicolumn{1}{r|}{DeepPrefetcher}& 50.4  & 50.4  & 50.5  & 50.5  & 50.5  & 50.5  & 50.5  & 50.6  & 50.6  & 50.6  & 50.7  & 50.8  & 51.0  & 51.0  & 51.1  & 51.1  & 51.1  & 51.7  & 51.7  & 51.7 \\
\multicolumn{1}{r|}{Delta-LSTM} & 56.3 & 56.4 & 56.4 & 56.4 & 56.4 & 56.5 & 56.5 & 56.5 & 56.6 & 56.6 & 56.8 & 57.1 & 57.1 & 57.1 & 57.2 & 57.3 & 57.8 & 57.8 & 57.8 & 57.9 \\ \hline
\multicolumn{1}{r|}{SGDP} & 76.0 & 76.0 & 76.1 & 76.1 & 76.1 & 76.1 & 76.2 & 76.4 & 76.5 & 76.5 & 76.6 & 76.8 & 76.8 & 76.9 & 77.3 & 77.3 & 77.4 & 77.4 & 77.4 & 77.5 \\
\multicolumn{1}{r|}{SGDP$_l$} & \textbf{78.5} & \textbf{78.5} & \textbf{78.5} & \textbf{78.6} & \textbf{78.6} & \textbf{78.6} & \textbf{78.6} & \textbf{78.9} & \textbf{79.0} & \textbf{79.0} & \textbf{79.0} & \textbf{79.2} & \textbf{79.3} & \textbf{79.3} & \textbf{79.3} & \textbf{79.7} & \textbf{79.7} & \textbf{79.7} & \textbf{79.7} & \textbf{79.8} \\
\multicolumn{1}{r|}{SGDP$_p$} & 48.0 & 48.1 & 48.1 & 48.1 & 48.1 & 48.1 & 48.2 & 48.2 & 48.2 & 48.3 & 48.3 & 48.6 & 48.9 & 48.9 & 49.0 & 49.3 & 49.6 & 49.6 & 49.6 & 49.6 \\ \hline
\multicolumn{21}{c}{EPR@N} \\ \hline
\multicolumn{1}{c|}{\diagbox{Methods}{Cache sizes}}  & \multicolumn{1}{c}{5} & \multicolumn{1}{c}{10} & \multicolumn{1}{c}{20} & \multicolumn{1}{c}{30} & \multicolumn{1}{c}{40} & \multicolumn{1}{c}{50} & \multicolumn{1}{c}{60} & \multicolumn{1}{c}{70} & \multicolumn{1}{c}{80} & \multicolumn{1}{c}{90} & \multicolumn{1}{c}{100} & \multicolumn{1}{c}{200} & \multicolumn{1}{c}{300} & \multicolumn{1}{c}{400} & \multicolumn{1}{c}{500} & \multicolumn{1}{c}{600} & \multicolumn{1}{c}{700} & \multicolumn{1}{c}{800} & \multicolumn{1}{c}{900} & \multicolumn{1}{c}{1000} \\ \hline
\multicolumn{1}{r|}{Naive} & 47.9 & 48.0 & 48.1 & 48.1 & 48.1 & 48.2 & 48.2 & 48.2 & 48.2 & 48.2 & 48.3 & 48.5 & 48.5 & 48.4 & 48.4 & 48.4 & 48.7 & 48.7 & 48.7 & 48.7 \\
\multicolumn{1}{r|}{Stride} & 81.6 & 81.6 & 81.7 & 81.7 & 81.7 & 81.8 & 81.8 & 81.8 & 81.9 & 82.0 & 82.0 & 82.1 & 82.1 & 82.2 & 82.2 & 82.2 & 82.3 & 82.3 & 82.3 & 82.3 \\
\multicolumn{1}{r|}{ARIMA} & 0.1 & 0.2 & 0.3 & 0.3 & 0.3 & 0.3 & 0.3 & 0.3 & 0.3 & 0.3 & 0.3 & 0.3 & 0.3 & 0.3 & 0.3 & 0.3 & 0.3 & 0.3 & 0.3 & 0.3 \\
\multicolumn{1}{r|}{Informer} & 0.0 & 0.0 & 0.0 & 0.0 & 0.0 & 0.0 & 0.0 & 0.0 & 0.0 & 0.0 & 0.0 & 0.0 & 0.0 & 0.0 & 0.0 & 0.0 & 0.0 & 0.0 & 0.0 & 0.0 \\
\multicolumn{1}{r|}{DeepPrefetcher}& 50.4  & 50.4  & 50.5  & 50.5  & 50.5  & 50.5  & 50.5  & 50.6  & 50.6  & 50.7  & 50.7  & 50.8  & 51.0  & 51.0  & 51.0  & 51.0  & 51.0  & 51.2  & 51.2  & 51.2 \\
\multicolumn{1}{r|}{Delta-LSTM} & 65.3 & 66.2 & 66.3 & 66.3 & 66.3 & 66.4 & 66.4 & 66.4 & 66.5 & 66.5 & 66.8 & 67.0 & 67.0 & 67.0 & 67.0 & 67.1 & 67.2 & 67.2 & 67.2 & 67.2 \\ \hline
\multicolumn{1}{r|}{SGDP} & \textbf{88.8} & \textbf{88.9} & \textbf{88.9} & \textbf{88.9} & \textbf{89.0} & \textbf{89.0} & \textbf{89.0} & \textbf{89.3} & \textbf{89.4} & \textbf{89.5} & \textbf{89.5} & \textbf{89.7} & \textbf{89.7} & \textbf{89.8} & \textbf{90.0} & \textbf{90.0} & \textbf{90.1} & \textbf{90.1} & \textbf{90.1} & \textbf{90.1} \\
\multicolumn{1}{r|}{SGDP$_l$} & 83.6 & 83.6 & 83.6 & 83.6 & 83.7 & 83.7 & 83.7 & 84.0 & 84.1 & 84.1 & 84.2 & 84.4 & 84.4 & 84.5 & 84.5 & 84.7 & 84.7 & 84.7 & 84.7 & 84.7 \\
\multicolumn{1}{r|}{SGDP$_p$} & 70.6 & 72.1 & 72.5 & 72.6 & 72.6 & 72.6 & 72.7 & 72.8 & 72.9 & 73.0 & 73.1 & 73.7 & 74.4 & 74.5 & 74.6 & 74.7 & 74.9 & 75.1 & 75.1 & 75.1 \\ \hline \hline 
\end{tabular}
        }
	\label{tab:hw_3}
\end{table*}

\begin{table*}[!h]
	\centering
	\caption{Results of single-step prefetching based on different cache sizes about dataset \textbf{hm\_1}.}
	\renewcommand{\arraystretch}{1.3}
	\resizebox{\textwidth}{!}{
\begin{tabular}{rllllllllllllllllllll}
\hline \hline
\multicolumn{21}{c}{HR@N} \\ \hline
\multicolumn{1}{c|}{\diagbox{Methods}{Cache sizes}}  & \multicolumn{1}{c}{5} & \multicolumn{1}{c}{10} & \multicolumn{1}{c}{20} & \multicolumn{1}{c}{30} & \multicolumn{1}{c}{40} & \multicolumn{1}{c}{50} & \multicolumn{1}{c}{60} & \multicolumn{1}{c}{70} & \multicolumn{1}{c}{80} & \multicolumn{1}{c}{90} & \multicolumn{1}{c}{100} & \multicolumn{1}{c}{200} & \multicolumn{1}{c}{300} & \multicolumn{1}{c}{400} & \multicolumn{1}{c}{500} & \multicolumn{1}{c}{600} & \multicolumn{1}{c}{700} & \multicolumn{1}{c}{800} & \multicolumn{1}{c}{900} & \multicolumn{1}{c}{1000} \\ \hline
\multicolumn{1}{r|}{No\_pre} & 1.0 & 2.7 & 5.5 & 8.5 & 11.9 & 14.3 & 16.6 & 19.4 & 21.5 & 23.5 & 25.3 & 42.2 & 68.7 & 91.9 & 95.4 & 96.2 & 96.8 & 97.3 & 97.9 & 98.3 \\
\multicolumn{1}{r|}{Naive} & 30.8 & 31.7 & 33.2 & 34.6 & 36.0 & 37.4 & 38.8 & 40.2 & 41.5 & 42.7 & 43.8 & 53.4 & 60.5 & 68.2 & 78.2 & 88.0 & 94.2 & 96.4 & 97.1 & 97.4 \\
\multicolumn{1}{r|}{Stride} & 25.7 & 27.1 & 29.6 & 32.1 & 34.9 & 37.1 & 39.3 & 41.8 & 43.7 & 45.3 & 47.0 & 59.8 & 75.6 & 95.3 & 97.3 & 98.0 & 98.4 & 98.6 & 98.9 & 99.1 \\
\multicolumn{1}{r|}{ARIMA} & 2.4 & 3.5 & 5.6 & 7.5 & 9.4 & 11.3 & 13.1 & 14.7 & 16.3 & 17.7 & 19.0 & 30.4 & 38.7 & 48.5 & 60.9 & 74.9 & 86.3 & 91.7 & 93.8 & 95.2 \\
\multicolumn{1}{r|}{Informer} & 0.5 & 1.1 & 2.8 & 4.1 & 5.7 & 7.2 & 8.5 & 10.0 & 11.5 & 12.7 & 14.0 & 24.1 & 31.4 & 38.6 & 48.1 & 59.3 & 71.1 & 81.1 & 86.8 & 90.4 \\
\multicolumn{1}{r|}{DeepPrefetcher} & 36.7  & 38.5  & 41.7  & 44.6  & 47.1  & 49.6  & 52.0  & 54.1  & 55.9  & 57.5  & 59.1  & 71.2  & 80.5  & 94.3  & 97.2  & 98.0  & 98.6  & 99.0  & 99.1  & 99.3 \\
\multicolumn{1}{r|}{Delta-LSTM} & 28.3 & 30.0 & 32.9 & 35.6 & 38.3 & 40.9 & 43.1 & 45.0 & 47.0 & 49.0 & 50.6 & 63.6 & 78.3 & 95.9 & 97.8 & 98.4 & 98.8 & 99.0 & 99.1 & 99.3 \\ \hline
\multicolumn{1}{r|}{SGDP} & 36.7 & 38.1 & 40.4 & 42.7 & 45.0 & 47.1 & 48.8 & 50.9 & 52.8 & 54.3 & 55.7 & 67.4 & \textbf{81.2} & \textbf{96.8} & \textbf{98.4} & \textbf{98.8} & \textbf{99.1} & \textbf{99.2} & \textbf{99.3} & \textbf{99.4} \\
\multicolumn{1}{r|}{SGDP$_l$} & 40.6 & 43.1 & 46.5 & 49.1 & 51.4 & 53.5 & 55.4 & 57.1 & 58.6 & 60.1 & 61.4 & 70.3 & 78.5 & 90.2 & 97.4 & 98.2 & 98.6 & 98.8 & 98.9 & 99.1 \\
\multicolumn{1}{r|}{SGDP$_p$} & \textbf{41.4} & \textbf{43.9} & \textbf{47.5} & \textbf{50.3} & \textbf{52.7} & \textbf{54.9} & \textbf{56.8} & \textbf{58.5} & \textbf{59.9} & \textbf{61.4} & \textbf{62.9} & \textbf{71.8} & 80.3 & 92.8 & 97.5 & 98.3 & 98.6 & 98.9 & 99.2 & 99.4 \\ \hline
\multicolumn{21}{c}{EPR@N} \\ \hline
\multicolumn{1}{c|}{\diagbox{Methods}{Cache sizes}}  & \multicolumn{1}{c}{5} & \multicolumn{1}{c}{10} & \multicolumn{1}{c}{20} & \multicolumn{1}{c}{30} & \multicolumn{1}{c}{40} & \multicolumn{1}{c}{50} & \multicolumn{1}{c}{60} & \multicolumn{1}{c}{70} & \multicolumn{1}{c}{80} & \multicolumn{1}{c}{90} & \multicolumn{1}{c}{100} & \multicolumn{1}{c}{200} & \multicolumn{1}{c}{300} & \multicolumn{1}{c}{400} & \multicolumn{1}{c}{500} & \multicolumn{1}{c}{600} & \multicolumn{1}{c}{700} & \multicolumn{1}{c}{800} & \multicolumn{1}{c}{900} & \multicolumn{1}{c}{1000} \\ \hline
\multicolumn{1}{r|}{Naive} & 30.5 & 30.5 & 30.8 & 30.9 & 31.0 & 31.0 & 31.1 & 31.1 & 31.2 & 31.2 & 31.2 & 31.2 & 30.0 & 26.9 & 21.8 & 16.6 & 11.8 & 7.5 & 6.2 & 5.6 \\
\multicolumn{1}{r|}{Stride} & 82.0 & 82.3 & 82.7 & 83.0 & 83.3 & 83.5 & 83.8 & 84.0 & 84.1 & 84.2 & 84.4 & 84.3 & 79.0 & 83.6 & 86.1 & 87.7 & 88.0 & 87.8 & \textbf{87.4} & \textbf{88.4} \\
\multicolumn{1}{r|}{ARIMA} & 2.3 & 2.7 & 3.2 & 3.6 & 4.0 & 4.2 & 4.5 & 4.6 & 4.8 & 5.0 & 5.2 & 6.3 & 6.9 & 6.5 & 5.6 & 4.0 & 3.0 & 2.5 & 2.3 & 2.5 \\
\multicolumn{1}{r|}{Informer} & 0.1 & 0.1 & 0.2 & 0.3 & 0.4 & 0.4 & 0.5 & 0.6 & 0.6 & 0.7 & 0.7 & 1.1 & 1.4 & 1.5 & 1.4 & 1.2 & 1.0 & 0.9 & 0.8 & 0.7 \\
\multicolumn{1}{r|}{DeepPrefetcher} & 37.0  & 38.5  & 41.5  & 44.0  & 46.2  & 48.3  & 50.1  & 51.9  & 53.5  & 54.8  & 56.0  & 65.3  & 67.1  & 53.5  & 48.4  & 41.5  & 42.1  & 45.2  & 47.4  & 46.1  \\
\multicolumn{1}{r|}{Delta-LSTM} & 55.9 & 57.7 & 60.9 & 63.4 & 65.4 & 67.1 & 68.6 & 69.8 & 70.8 & 71.9 & 72.8 & 79.0 & 79.6 & 78.1 & 81.8 & 83.4 & 85.7 & 87.0 & 86.6 & 87.6 \\ \hline
\multicolumn{1}{r|}{SGDP} & \textbf{86.5} & \textbf{87.8} & \textbf{88.7} & \textbf{89.1} & \textbf{89.2} & \textbf{89.5} & \textbf{89.7} & \textbf{89.8} & \textbf{89.9} & \textbf{90.0} & \textbf{90.1} & \textbf{90.9} & \textbf{87.8} &\textbf{ 86.7} & \textbf{88.1} & \textbf{89.1} & \textbf{88.9} & \textbf{88.7} & 87.3 & 86.2 \\
\multicolumn{1}{r|}{SGDP$_l$} & 43.3 & 46.3 & 50.3 & 53.0 & 54.8 & 56.3 & 57.6 & 58.7 & 59.6 & 60.2 & 60.8 & 63.7 & 61.5 & 49.0 & 33.9 & 27.1 & 26.3 & 26.6 & 26.7 & 24.4 \\
\multicolumn{1}{r|}{SGDP$_p$} & 43.4 & 46.8 & 51.2 & 54.3 & 56.5 & 58.3 & 59.9 & 61.1 & 62.1 & 63.0 & 63.8 & 67.7 & 65.9 & 51.4 & 41.6 & 35.5 & 35.1 & 35.5 & 36.3 & 34.8 \\ \hline \hline
\end{tabular}
        }
	\label{tab:hm_1}
\end{table*}

\begin{table*}[!h]
	\centering
	\caption{Results of single-step prefetching based on different cache sizes about dataset \textbf{mds\_0}.}
	\renewcommand{\arraystretch}{1.3}
	\resizebox{\textwidth}{!}{

\begin{tabular}{rllllllllllllllllllll}
\hline \hline
\multicolumn{21}{c}{HR@N} \\ \hline
\multicolumn{1}{c|}{\diagbox{Methods}{Cache sizes}} & \multicolumn{1}{c}{5} & \multicolumn{1}{c}{10} & \multicolumn{1}{c}{20} & \multicolumn{1}{c}{30} & \multicolumn{1}{c}{40} & \multicolumn{1}{c}{50} & \multicolumn{1}{c}{60} & \multicolumn{1}{c}{70} & \multicolumn{1}{c}{80} & \multicolumn{1}{c}{90} & \multicolumn{1}{c}{100} & \multicolumn{1}{c}{200} & \multicolumn{1}{c}{300} & \multicolumn{1}{c}{400} & \multicolumn{1}{c}{500} & \multicolumn{1}{c}{600} & \multicolumn{1}{c}{700} & \multicolumn{1}{c}{800} & \multicolumn{1}{c}{900} & \multicolumn{1}{c}{1000} \\ \hline
\multicolumn{1}{r|}{No\_pre} & 10.1 & 13.2 & 18.1 & 21.0 & 24.2 & 26.7 & 30.0 & 32.4 & 33.4 & 34.2 & 35.0 & 45.8 & 49.6 & 51.3 & 52.6 & 55.0 & 57.9 & 58.5 & 59.3 & 61.0 \\
\multicolumn{1}{r|}{Naive} & 50.0 & 54.3 & 58.4 & 61.2 & 63.1 & 64.6 & 65.8 & 66.8 & 67.5 & 67.9 & 68.2 & 71.1 & 75.0 & 78.9 & 81.1 & 82.9 & 83.9 & 84.5 & 84.9 & 85.2 \\
\multicolumn{1}{r|}{Stride} & 44.0 & 47.3 & 51.8 & 54.2 & 55.8 & 57.5 & 59.0 & 60.3 & 60.9 & 61.5 & 62.2 & 72.0 & 75.6 & 76.9 & 77.4 & 77.8 & 78.5 & 78.9 & 79.2 & 79.8 \\
\multicolumn{1}{r|}{ARIMA} & 13.2 & 16.6 & 20.7 & 24.0 & 27.4 & 30.2 & 32.6 & 35.0 & 36.1 & 36.9 & 37.4 & 41.2 & 46.1 & 50.3 & 52.8 & 54.3 & 55.2 & 56.1 & 57.6 & 58.3 \\
\multicolumn{1}{r|}{Informer} & 6.2 & 9.6 & 13.1 & 16.0 & 18.6 & 20.4 & 22.1 & 23.7 & 25.4 & 26.9 & 28.3 & 35.9 & 40.2 & 44.1 & 47.2 & 49.0 & 50.9 & 52.3 & 53.5 & 54.5 \\
\multicolumn{1}{r|}{DeepPrefetcher} & 55.7  & 60.7  & 65.3  & 68.1  & 69.8  & 71.1  & 71.9  & 72.5  & 73.0  & 73.4  & 73.7  & 78.9  & 84.4  & 86.2  & 87.1  & 87.4  & 87.7  & 87.9  & 88.1  & 88.5  \\
\multicolumn{1}{r|}{Delta-LSTM} & 53.8 & 57.3 & 61.5 & 63.6 & 65.0 & 66.1 & 67.1 & 67.9 & 68.5 & 69.0 & 69.6 & 78.6 & 82.4 & 83.9 & 84.3 & 84.7 & 85.0 & 85.3 & 85.7 & 86.2 \\ \hline
\multicolumn{1}{r|}{SGDP} & 62.5 & 66.0 & 69.7 & 71.7 & 72.8 & 73.7 & 74.4 & 74.9 & 75.4 & 75.8 & 76.3 & 84.5 & 88.4 & 89.9 & 90.3 & 90.5 & 90.7 & 91.0 & 91.3 & 91.6 \\
\multicolumn{1}{r|}{SGDP$_l$} & 62.3 & 66.1 & 70.1 & 72.4 & 73.9 & 75.1 & 76.0 & 76.5 & 76.9 & 77.2 & 77.5 & 82.0 & 87.5 & 89.7 & 90.6 & 91.1 & 91.4 & 91.6 & 91.8 & 92.1 \\
\multicolumn{1}{r|}{SGDP$_p$} & \textbf{63.5} & \textbf{67.4} & \textbf{71.8} & \textbf{74.4} & \textbf{76.0} & \textbf{77.2} & \textbf{78.0} & \textbf{78.6} & \textbf{79.1} & \textbf{79.4} & \textbf{79.8} & \textbf{86.1} & \textbf{89.9} & \textbf{91.2} & \textbf{91.6} & \textbf{91.8} & \textbf{92.0} & \textbf{92.2} & \textbf{92.4} & \textbf{92.6} \\ \hline
\multicolumn{21}{c}{EPR@N} \\ \hline
\multicolumn{1}{c|}{\diagbox{Methods}{Cache sizes}} & \multicolumn{1}{c}{5} & \multicolumn{1}{c}{10} & \multicolumn{1}{c}{20} & \multicolumn{1}{c}{30} & \multicolumn{1}{c}{40} & \multicolumn{1}{c}{50} & \multicolumn{1}{c}{60} & \multicolumn{1}{c}{70} & \multicolumn{1}{c}{80} & \multicolumn{1}{c}{90} & \multicolumn{1}{c}{100} & \multicolumn{1}{c}{200} & \multicolumn{1}{c}{300} & \multicolumn{1}{c}{400} & \multicolumn{1}{c}{500} & \multicolumn{1}{c}{600} & \multicolumn{1}{c}{700} & \multicolumn{1}{c}{800} & \multicolumn{1}{c}{900} & \multicolumn{1}{c}{1000} \\ \hline
\multicolumn{1}{r|}{Naive} & 45.8 & 47.8 & 50.4 & 51.8 & 52.6 & 52.5 & 51.9 & 51.0 & 50.9 & 51.0 & 51.1 & 52.1 & 52.8 & 53.1 & 52.9 & 52.8 & 50.8 & 51.2 & 51.8 & 52.2 \\
\multicolumn{1}{r|}{Stride} & \textbf{79.2} & \textbf{82.3} & \textbf{86.9} & \textbf{89.1} & \textbf{90.1} & \textbf{90.5} & \textbf{90.5} & \textbf{90.5} & \textbf{90.6} & \textbf{90.6} & \textbf{90.6} & \textbf{90.7} & \textbf{90.8} & \textbf{90.9} & \textbf{90.7} & \textbf{90.6} & \textbf{89.9} & \textbf{89.8} & \textbf{89.8} & \textbf{89.8} \\
\multicolumn{1}{r|}{ARIMA} & 8.1 & 8.6 & 9.5 & 9.9 & 10.0 & 10.1 & 9.9 & 9.3 & 9.3 & 9.2 & 9.2 & 9.6 & 9.8 & 10.2 & 10.4 & 10.7 & 11.0 & 11.3 & 11.7 & 12.0 \\
\multicolumn{1}{r|}{Informer} & 0.2 & 0.3 & 0.5 & 0.8 & 1.0 & 1.0 & 1.1 & 1.1 & 1.1 & 1.1 & 1.2 & 2.1 & 3.0 & 3.7 & 4.2 & 4.5 & 4.7 & 4.8 & 5.0 & 5.2 \\
\multicolumn{1}{r|}{DeepPrefetcher} & 63.1  & 66.9  & 71.3  & 73.9  & 75.4  & 76.1  & 76.2  & 76.3  & 76.7  & 77.2  & 77.5  & 80.2  & 82.3  & 83.1  & 83.5  & 83.7  & 83.2  & 83.2  & 83.2  & 83.3 \\
\multicolumn{1}{r|}{Delta-LSTM} & 77.0 & 80.2 & 84.2 & 86.1 & 87.0 & 87.3 & 87.4 & 87.4 & 87.6 & 87.6 & 87.8 & 89.1 & 90.0 & 90.4 & 90.4 & 90.1 & 89.7 & 89.6 & 89.8 & 89.8 \\ \hline
\multicolumn{1}{r|}{SGDP} & 77.2 & 80.2 & 83.9 & 85.7 & 86.5 & 86.8 & 86.8 & 86.6 & 86.9 & 87.0 & 87.0 & 87.9 & 88.5 & 88.8 & 88.9 & 88.9 & 88.4 & 88.4 & 88.5 & 88.4 \\
\multicolumn{1}{r|}{SGDP$_l$} & 62.4 & 65.4 & 69.2 & 71.3 & 72.4 & 72.7 & 72.8 & 72.7 & 73.2 & 73.5 & 73.9 & 76.5 & 78.3 & 79.2 & 79.8 & 80.0 & 79.2 & 79.4 & 79.6 & 79.6 \\
\multicolumn{1}{r|}{SGDP$_p$} & 64.9 & 68.7 & 73.8 & 77.0 & 78.9 & 79.6 & 80.0 & 80.4 & 81.0 & 81.4 & 81.9 & 85.3 & 87.5 & 88.1 & 88.4 & 88.4 & 87.8 & 87.8 & 88.0 & 87.9 \\ \hline \hline
\end{tabular}

        }
	\label{tab:mds_0}
\end{table*}

\begin{table*}[!h]
	\centering
	\caption{Results of single-step prefetching based on different cache sizes about dataset \textbf{proj\_0}.}
	\renewcommand{\arraystretch}{1.3}
	\resizebox{\textwidth}{!}{

\begin{tabular}{rllllllllllllllllllll}
\hline \hline
\multicolumn{21}{c}{HR@N} \\ \hline
\multicolumn{1}{c|}{\diagbox{Methods}{Cache sizes}} & \multicolumn{1}{c}{5} & \multicolumn{1}{c}{10} & \multicolumn{1}{c}{20} & \multicolumn{1}{c}{30} & \multicolumn{1}{c}{40} & \multicolumn{1}{c}{50} & \multicolumn{1}{c}{60} & \multicolumn{1}{c}{70} & \multicolumn{1}{c}{80} & \multicolumn{1}{c}{90} & \multicolumn{1}{c}{100} & \multicolumn{1}{c}{200} & \multicolumn{1}{c}{300} & \multicolumn{1}{c}{400} & \multicolumn{1}{c}{500} & \multicolumn{1}{c}{600} & \multicolumn{1}{c}{700} & \multicolumn{1}{c}{800} & \multicolumn{1}{c}{900} & \multicolumn{1}{c}{1000} \\ \hline
\multicolumn{1}{r|}{No\_pre} & 4.1 & 6.1 & 8.9 & 11.0 & 14.1 & 17.4 & 23.1 & 26.3 & 27.3 & 28.2 & 28.7 & 30.8 & 32.3 & 32.8 & 33.2 & 33.5 & 34.0 & 34.8 & 35.0 & 35.2 \\
\multicolumn{1}{r|}{Naive} & 59.0 & 61.1 & 63.6 & 65.1 & 66.3 & 67.2 & 68.2 & 69.2 & 69.6 & 69.9 & 70.1 & 71.1 & 71.8 & 72.5 & 73.0 & 73.4 & 73.7 & 73.9 & 74.2 & 74.3 \\
\multicolumn{1}{r|}{Stride} & 48.7 & 51.0 & 53.1 & 54.5 & 55.7 & 57.0 & 58.9 & 60.1 & 60.5 & 60.8 & 61.1 & 62.5 & 63.8 & 64.1 & 64.3 & 64.7 & 64.9 & 65.2 & 65.3 & 65.4 \\
\multicolumn{1}{r|}{ARIMA} & 10.7 & 12.9 & 15.6 & 18.1 & 20.8 & 23.6 & 27.5 & 30.8 & 32.1 & 33.0 & 33.5 & 35.2 & 36.1 & 37.0 & 37.7 & 38.3 & 38.7 & 38.9 & 39.0 & 39.3 \\
\multicolumn{1}{r|}{Informer} & 2.3 & 3.9 & 6.0 & 7.7 & 9.2 & 10.6 & 12.1 & 13.7 & 15.4 & 17.3 & 19.8 & 29.6 & 30.8 & 31.9 & 32.6 & 33.4 & 33.8 & 34.2 & 34.5 & 34.7 \\
\multicolumn{1}{r|}{DeepPrefetcher} & 70.4  & 72.6  & 74.7  & 75.9  & 76.8  & 77.4  & 78.1  & 78.5  & 78.7  & 78.9  & 79.1  & 80.2  & 81.1  & 81.8  & 81.9  & 82.1  & 82.2  & 82.5  & 82.7  & 82.8 \\
\multicolumn{1}{r|}{Delta-LSTM} & 60.5 & 62.3 & 64.0 & 65.0 & 65.9 & 66.7 & 67.7 & 68.4 & 68.7 & 68.9 & 69.1 & 70.5 & 71.7 & 72.1 & 72.3 & 72.6 & 72.8 & 73.1 & 73.3 & 73.3 \\ \hline
\multicolumn{1}{r|}{SGDP} & 71.5 & 73.4 & 75.0 & 75.9 & 76.5 & 77.1 & 77.6 & 78.0 & 78.2 & 78.4 & 78.5 & 79.7 & 80.7 & 81.1 & 81.2 & 81.5 & 81.7 & 81.9 & 82.1 & 82.1 \\
\multicolumn{1}{r|}{SGDP$_l$} & \textbf{73.5} & \textbf{75.5} & \textbf{77.3} & \textbf{78.3} & \textbf{79.0} & \textbf{79.6} & \textbf{80.2} & \textbf{80.6} & \textbf{80.8} & 81.0 & 81.1 & 82.0 & 82.9 & 83.6 & 83.8 & 84.0 & 84.2 & 84.3 & 84.5 & 84.6 \\
\multicolumn{1}{r|}{SGDP$_p$} & 71.5 & 73.7 & 76.2 & 77.5 & 78.4 & 79.2 & 80.0 & 80.5 & 80.8 & \textbf{81.0} & \textbf{81.3} & \textbf{82.6} & \textbf{83.6} & \textbf{84.2} & \textbf{84.4} & \textbf{84.5} & \textbf{84.7} & \textbf{84.9} & \textbf{85.1} & \textbf{85.2} \\ \hline
\multicolumn{21}{c}{EPR@N} \\ \hline
\multicolumn{1}{c|}{\diagbox{Methods}{Cache sizes}} & \multicolumn{1}{c}{5} & \multicolumn{1}{c}{10} & \multicolumn{1}{c}{20} & \multicolumn{1}{c}{30} & \multicolumn{1}{c}{40} & \multicolumn{1}{c}{50} & \multicolumn{1}{c}{60} & \multicolumn{1}{c}{70} & \multicolumn{1}{c}{80} & \multicolumn{1}{c}{90} & \multicolumn{1}{c}{100} & \multicolumn{1}{c}{200} & \multicolumn{1}{c}{300} & \multicolumn{1}{c}{400} & \multicolumn{1}{c}{500} & \multicolumn{1}{c}{600} & \multicolumn{1}{c}{700} & \multicolumn{1}{c}{800} & \multicolumn{1}{c}{900} & \multicolumn{1}{c}{1000} \\ \hline
\multicolumn{1}{r|}{Naive} & 57.6 & 58.7 & 61.0 & 62.3 & 62.9 & 62.7 & 61.9 & 60.1 & 59.8 & 59.8 & 59.7 & 60.2 & 60.4 & 60.6 & 60.7 & 60.9 & 60.9 & 60.9 & 60.8 & 60.8 \\
\multicolumn{1}{r|}{Stride} & 80.3 & 82.5 & 85.6 & 87.3 & 88.4 & \textbf{88.7} & \textbf{88.4} & \textbf{88.0} & \textbf{88.1} & \textbf{88.1} & \textbf{88.1} & \textbf{88.3} & \textbf{88.3} & \textbf{88.3} & \textbf{88.3} & \textbf{88.4} & \textbf{88.4} & \textbf{88.3} & \textbf{88.3} & \textbf{88.3} \\
\multicolumn{1}{r|}{ARIMA} & 11.6 & 12.0 & 12.6 & 13.0 & 13.1 & 12.6 & 11.7 & 10.8 & 10.4 & 10.2 & 10.1 & 10.0 & 10.1 & 10.1 & 10.2 & 10.2 & 10.3 & 10.3 & 10.3 & 10.4 \\
\multicolumn{1}{r|}{Informer} & 0.0 & 0.1 & 0.1 & 0.2 & 0.3 & 0.3 & 0.4 & 0.4 & 0.5 & 0.5 & 0.5 & 1.0 & 1.4 & 1.8 & 2.0 & 2.2 & 2.2 & 2.3 & 2.3 & 2.3 \\
\multicolumn{1}{r|}{DeepPrefetcher}& 73.0  & 75.0  & 77.5  & 78.9  & 79.7  & 79.6  & 78.9  & 78.3  & 78.4  & 78.5  & 78.6  & 79.7  & 80.3  & 80.8  & 81.0  & 81.1  & 81.2  & 81.3  & 81.4  & 81.5  \\
\multicolumn{1}{r|}{Delta-LSTM} & \textbf{82.8} & \textbf{84.3} & 86.1 & 87.1 & 87.5 & 87.4 & 86.7 & 86.2 & 86.1 & 86.1 & 86.2 & 86.6 & 86.9 & 87.1 & 87.2 & 87.3 & 87.3 & 87.4 & 87.3 & 87.4 \\ \hline
\multicolumn{1}{r|}{SGDP} & 82.2 & 84.0 & \textbf{86.3} & \textbf{87.6} & \textbf{88.4} & 88.4 & 88.0 & 87.6 & 87.6 & 87.6 & 87.6 & 87.9 & 88.0 & 88.1 & 88.1 & 88.2 & 88.2 & 88.2 & 88.2 & 88.2 \\
\multicolumn{1}{r|}{SGDP$_l$} & 77.8 & 79.8 & 82.2 & 83.5 & 84.3 & 84.3 & 83.8 & 83.4 & 83.4 & 83.5 & 83.6 & 84.5 & 84.9 & 85.3 & 85.4 & 85.5 & 85.6 & 85.6 & 85.6 & 85.6 \\
\multicolumn{1}{r|}{SGDP$_p$} & 72.3 & 74.5 & 77.5 & 79.2 & 80.3 & 80.5 & 80.0 & 79.6 & 79.8 & 79.9 & 80.2 & 81.6 & 82.5 & 83.0 & 83.3 & 83.5 & 83.6 & 83.7 & 83.8 & 83.9 \\ \hline \hline
\end{tabular}

        }
	\label{tab:proj_0}
\end{table*}

\begin{table*}[!h]
	\centering
	\caption{Results of single-step prefetching based on different cache sizes about dataset \textbf{prxy\_0}.}
	\renewcommand{\arraystretch}{1.3}
	\resizebox{\textwidth}{!}{

\begin{tabular}{rllllllllllllllllllll}
\hline \hline
\multicolumn{21}{c}{HR@N} \\ \hline
\multicolumn{1}{c|}{\diagbox{Methods}{Cache sizes}} & \multicolumn{1}{c}{5} & \multicolumn{1}{c}{10} & \multicolumn{1}{c}{20} & \multicolumn{1}{c}{30} & \multicolumn{1}{c}{40} & \multicolumn{1}{c}{50} & \multicolumn{1}{c}{60} & \multicolumn{1}{c}{70} & \multicolumn{1}{c}{80} & \multicolumn{1}{c}{90} & \multicolumn{1}{c}{100} & \multicolumn{1}{c}{200} & \multicolumn{1}{c}{300} & \multicolumn{1}{c}{400} & \multicolumn{1}{c}{500} & \multicolumn{1}{c}{600} & \multicolumn{1}{c}{700} & \multicolumn{1}{c}{800} & \multicolumn{1}{c}{900} & \multicolumn{1}{c}{1000} \\ \hline
\multicolumn{1}{r|}{No\_pre} & 14.9 & 20.1 & 25.1 & 27.5 & 30.0 & 32.4 & 35.7 & 38.0 & 39.2 & 40.1 & 40.7 & 45.9 & 47.5 & 47.8 & 48.0 & 48.4 & 48.5 & 48.6 & 48.7 & 48.8 \\
\multicolumn{1}{r|}{Naive} & 40.4 & 46.4 & 52.4 & 56.0 & 58.2 & 59.9 & 61.3 & 62.5 & 63.3 & 63.9 & 64.3 & 66.5 & 67.9 & 69.5 & 70.5 & 71.3 & 71.9 & 72.3 & 72.5 & 72.7 \\
\multicolumn{1}{r|}{Stride} & 34.5 & 40.3 & 45.1 & 47.6 & 49.4 & 51.3 & 53.2 & 54.6 & 55.3 & 55.9 & 56.5 & 61.1 & 62.6 & 62.9 & 63.2 & 63.5 & 63.6 & 63.7 & 63.8 & 63.8 \\
\multicolumn{1}{r|}{ARIMA} & 14.6 & 19.9 & 25.7 & 29.2 & 32.0 & 34.7 & 37.3 & 39.7 & 40.8 & 41.7 & 42.2 & 45.5 & 47.4 & 49.4 & 50.4 & 51.0 & 51.7 & 52.0 & 52.2 & 52.3 \\
\multicolumn{1}{r|}{Informer} & 8.4 & 13.7 & 18.6 & 21.8 & 24.0 & 25.7 & 26.9 & 28.2 & 29.7 & 30.8 & 32.1 & 39.8 & 41.9 & 43.1 & 44.0 & 44.9 & 45.5 & 46.1 & 46.6 & 46.9 \\
\multicolumn{1}{r|}{DeepPrefetcher}& 50.9  & 57.0  & 62.3  & 64.7  & 66.2  & 67.4  & 68.3  & 69.0  & 69.4  & 69.9  & 70.2  & 73.6  & 75.6  & 76.5  & 76.7  & 76.9  & 77.1  & 77.2  & 77.3  & 77.4  \\
\multicolumn{1}{r|}{Delta-LSTM} & 46.8 & 52.2 & 56.5 & 58.5 & 59.7 & 60.9 & 62.0 & 62.8 & 63.3 & 63.7 & 64.2 & 68.5 & 70.1 & 70.4 & 70.7 & 71.0 & 71.0 & 71.1 & 71.2 & 71.3 \\ \hline
\multicolumn{1}{r|}{SGDP} & 56.7 & 62.2 & 66.4 & 68.3 & 69.5 & 70.6 & 71.3 & 71.9 & 72.4 & 72.8 & 73.2 & 77.2 & 78.7 & 79.0 & 79.3 & 79.6 & 79.6 & 79.7 & 79.8 & 79.9 \\
\multicolumn{1}{r|}{SGDP$_l$} & \textbf{58.7} & \textbf{64.1} & \textbf{68.9} & \textbf{71.2} & \textbf{72.7} & \textbf{73.9} & \textbf{74.7} & \textbf{75.3} & \textbf{75.8} & \textbf{76.2} & \textbf{76.5} & 79.4 & 81.4 & 82.2 & 82.5 & \textbf{82.7} & \textbf{82.8} & \textbf{82.9} & \textbf{82.9} & \textbf{83.0} \\
\multicolumn{1}{r|}{SGDP$_p$} & 58.2 & 63.9 & 68.8 & 71.0 & 72.5 & 73.6 & 74.4 & 75.0 & 75.4 & 75.8 & 76.2 & \textbf{80.0} & \textbf{81.8} & \textbf{82.2} & \textbf{82.5} & 82.6 & 82.8 & 82.9 & 83.0 & 83.0 \\ \hline
\multicolumn{21}{c}{EPR@N} \\ \hline
\multicolumn{1}{c|}{\diagbox{Methods}{Cache sizes}} & \multicolumn{1}{c}{5} & \multicolumn{1}{c}{10} & \multicolumn{1}{c}{20} & \multicolumn{1}{c}{30} & \multicolumn{1}{c}{40} & \multicolumn{1}{c}{50} & \multicolumn{1}{c}{60} & \multicolumn{1}{c}{70} & \multicolumn{1}{c}{80} & \multicolumn{1}{c}{90} & \multicolumn{1}{c}{100} & \multicolumn{1}{c}{200} & \multicolumn{1}{c}{300} & \multicolumn{1}{c}{400} & \multicolumn{1}{c}{500} & \multicolumn{1}{c}{600} & \multicolumn{1}{c}{700} & \multicolumn{1}{c}{800} & \multicolumn{1}{c}{900} & \multicolumn{1}{c}{1000} \\ \hline
\multicolumn{1}{r|}{Naive} & 33.5 & 35.1 & 37.6 & 39.1 & 40.1 & 40.4 & 39.9 & 39.0 & 38.6 & 38.6 & 38.5 & 38.8 & 39.2 & 39.6 & 39.8 & 40.1 & 40.3 & 40.5 & 40.7 & 40.9 \\
\multicolumn{1}{r|}{Stride} & 65.2 & 69.6 & 75.3 & 78.9 & 81.2 & 81.9 & 81.7 & 81.4 & 81.3 & 81.2 & 81.1 & 81.1 & 81.2 & 81.3 & 81.3 & 81.3 & 81.4 & 81.4 & 81.4 & 81.4 \\
\multicolumn{1}{r|}{ARIMA} & 6.1 & 6.5 & 7.1 & 7.4 & 7.8 & 7.9 & 7.7 & 7.4 & 7.3 & 7.2 & 7.2 & 7.4 & 7.5 & 7.7 & 7.7 & 7.8 & 8.0 & 8.0 & 8.1 & 8.2 \\
\multicolumn{1}{r|}{Informer} & 0.0 & 0.0 & 0.0 & 0.0 & 0.0 & 0.0 & 0.0 & 0.0 & 0.0 & 0.0 & 0.0 & 0.0 & 0.1 & 0.1 & 0.1 & 0.1 & 0.1 & 0.1 & 0.2 & 0.2 \\
\multicolumn{1}{r|}{DeepPrefetcher}& 59.6  & 63.5  & 67.7  & 70.0  & 71.2  & 71.3  & 70.9  & 70.3  & 70.3  & 70.3  & 70.4  & 71.5  & 72.5  & 72.9  & 73.1  & 73.3  & 73.4  & 73.7  & 73.8  & 73.9  \\
\multicolumn{1}{r|}{Delta-LSTM} & 72.9 & 75.7 & 78.7 & 80.4 & 81.1 & 81.0 & 80.3 & 79.7 & 79.5 & 79.4 & 79.3 & 79.8 & 80.1 & 80.3 & 80.5 & 80.6 & 80.7 & 80.8 & 80.8 & 80.9 \\ \hline
\multicolumn{1}{r|}{SGDP} & \textbf{73.0} & \textbf{76.3} & \textbf{80.2} & \textbf{82.4} & \textbf{83.6} & \textbf{84.0} & \textbf{83.7} & \textbf{83.4} & \textbf{83.4} & \textbf{83.4} & \textbf{83.3} & \textbf{83.5} & \textbf{83.6} & \textbf{83.7} & \textbf{83.8} & \textbf{83.9} & \textbf{83.9} & \textbf{84.0} & \textbf{84.0} & \textbf{84.1} \\
\multicolumn{1}{r|}{SGDP$_l$} & 62.4 & 65.7 & 69.5 & 71.9 & 73.3 & 73.7 & 73.6 & 73.5 & 73.8 & 74.0 & 74.3 & 75.9 & 77.0 & 77.5 & 77.7 & 77.9 & 78.1 & 78.3 & 78.4 & 78.4 \\
\multicolumn{1}{r|}{SGDP$_p$} & 61.2 & 64.9 & 69.1 & 71.4 & 72.7 & 73.0 & 72.6 & 72.3 & 72.3 & 72.4 & 72.6 & 73.5 & 74.4 & 74.8 & 75.1 & 75.4 & 75.7 & 75.8 & 76.0 & 76.1 \\ \hline \hline
\end{tabular}

        }
	\label{tab:prxy_0}
\end{table*}

\begin{table*}[!h]
	\centering
	\caption{Results of single-step prefetching based on different cache sizes about dataset \textbf{src1\_2}.}
	\renewcommand{\arraystretch}{1.3}
	\resizebox{\textwidth}{!}{

\begin{tabular}{rllllllllllllllllllll}
\hline \hline
\multicolumn{21}{c}{HR@N} \\ \hline
\multicolumn{1}{c|}{\diagbox{Methods}{Cache sizes}}  & \multicolumn{1}{c}{5} & \multicolumn{1}{c}{10} & \multicolumn{1}{c}{20} & \multicolumn{1}{c}{30} & \multicolumn{1}{c}{40} & \multicolumn{1}{c}{50} & \multicolumn{1}{c}{60} & \multicolumn{1}{c}{70} & \multicolumn{1}{c}{80} & \multicolumn{1}{c}{90} & \multicolumn{1}{c}{100} & \multicolumn{1}{c}{200} & \multicolumn{1}{c}{300} & \multicolumn{1}{c}{400} & \multicolumn{1}{c}{500} & \multicolumn{1}{c}{600} & \multicolumn{1}{c}{700} & \multicolumn{1}{c}{800} & \multicolumn{1}{c}{900} & \multicolumn{1}{c}{1000} \\ \hline
\multicolumn{1}{r|}{No\_pre} & 1.6 & 3.9 & 7.6 & 11.3 & 15.6 & 20.5 & 27.6 & 31.5 & 32.8 & 33.9 & 34.8 & 39.5 & 42.6 & 44.6 & 45.7 & 46.3 & 47.0 & 47.4 & 47.9 & 48.2 \\
\multicolumn{1}{r|}{Naive} & 58.5 & 60.5 & 63.1 & 65.2 & 66.8 & 68.3 & 69.9 & 71.4 & 72.1 & 72.6 & 73.0 & 74.7 & 75.9 & 77.3 & 78.5 & 79.4 & 79.9 & 80.3 & 80.6 & 80.8 \\
\multicolumn{1}{r|}{Stride} & 45.9 & 48.3 & 51.0 & 53.2 & 55.3 & 57.6 & 60.3 & 62.0 & 62.7 & 63.3 & 63.8 & 66.2 & 69.2 & 70.9 & 71.7 & 72.1 & 72.6 & 72.9 & 73.2 & 73.4 \\
\multicolumn{1}{r|}{ARIMA} & 12.5 & 14.6 & 18.1 & 21.6 & 25.1 & 29.2 & 34.5 & 38.2 & 39.9 & 41.1 & 42.0 & 45.3 & 46.2 & 48.9 & 50.2 & 51.9 & 53.0 & 53.8 & 54.4 & 54.8 \\
\multicolumn{1}{r|}{Informer} & 0.7 & 1.7 & 4.1 & 6.2 & 8.0 & 10.1 & 12.3 & 14.4 & 16.5 & 19.3 & 22.5 & 34.8 & 36.7 & 39.3 & 40.8 & 42.4 & 43.5 & 44.0 & 44.7 & 45.3 \\
\multicolumn{1}{r|}{DeepPrefetcher} & 72.5  & 74.5  & 76.5  & 78.0  & 79.1  & 80.2  & 81.3  & 82.0  & 82.3  & 82.6  & 82.9  & 84.0  & 85.3  & 86.3  & 87.4  & 88.0  & 88.3  & 88.5  & 88.8  & 89.0  \\
\multicolumn{1}{r|}{Delta-LSTM} & 66.9 & 68.7 & 70.8 & 72.4 & 73.7 & 75.1 & 76.6 & 77.5 & 78.0 & 78.3 & 78.7 & 79.8 & 81.5 & 82.6 & 83.7 & 84.3 & 84.6 & 84.9 & 85.2 & 85.4 \\ \hline
\multicolumn{1}{r|}{SGDP} & 73.5 & 75.4 & 77.4 & 78.7 & 79.8 & 80.8 & 81.7 & 82.2 & 82.6 & 82.8 & 83.1 & 84.3 & 85.9 & 86.9 & 87.6 & 87.9 & 88.1 & 88.4 & 88.6 & 88.8 \\
\multicolumn{1}{r|}{SGDP$_l$} & \textbf{74.3} & \textbf{76.3} & \textbf{78.3} & \textbf{79.6} & \textbf{80.7} & 81.6 & 82.6 & 83.1 & 83.4 & 83.7 & 83.9 & 85.0 & 86.5 & 87.4 & 88.2 & 88.5 & 88.7 & 89.0 & 89.2 & 89.4 \\
\multicolumn{1}{r|}{SGDP$_p$} & 72.7 & 74.9 & 77.3 & 79.2 & 80.6 & \textbf{81.9} & \textbf{83.1} & \textbf{83.8} & \textbf{84.2} & \textbf{84.5} & \textbf{84.8} & \textbf{85.9} & \textbf{87.1} & \textbf{87.9} & \textbf{88.6} & \textbf{88.9} & \textbf{89.2} & \textbf{89.3} & \textbf{89.6} & \textbf{89.8} \\ \hline
\multicolumn{21}{c}{EPR@N} \\ \hline
\multicolumn{1}{c|}{\diagbox{Methods}{Cache sizes}}  & \multicolumn{1}{c}{5} & \multicolumn{1}{c}{10} & \multicolumn{1}{c}{20} & \multicolumn{1}{c}{30} & \multicolumn{1}{c}{40} & \multicolumn{1}{c}{50} & \multicolumn{1}{c}{60} & \multicolumn{1}{c}{70} & \multicolumn{1}{c}{80} & \multicolumn{1}{c}{90} & \multicolumn{1}{c}{100} & \multicolumn{1}{c}{200} & \multicolumn{1}{c}{300} & \multicolumn{1}{c}{400} & \multicolumn{1}{c}{500} & \multicolumn{1}{c}{600} & \multicolumn{1}{c}{700} & \multicolumn{1}{c}{800} & \multicolumn{1}{c}{900} & \multicolumn{1}{c}{1000} \\ \hline
\multicolumn{1}{r|}{Naive} & 58.3 & 59.9 & 63.3 & 65.2 & 65.9 & 65.8 & 64.9 & 63.2 & 63.0 & 63.0 & 63.1 & 63.8 & 64.3 & 64.8 & 65.1 & 65.3 & 65.6 & 65.8 & 66.1 & 66.3 \\
\multicolumn{1}{r|}{Stride} & 78.0 & 81.0 & \textbf{86.1} & \textbf{88.6} & \textbf{89.7} & \textbf{90.1} & \textbf{89.8} & \textbf{89.6} & \textbf{89.6} & \textbf{89.6} & \textbf{89.6} & \textbf{90.0} & \textbf{91.1} & \textbf{91.6} & \textbf{91.7} & \textbf{91.8} & \textbf{91.9} & \textbf{91.9} & \textbf{92.0} & \textbf{92.0} \\
\multicolumn{1}{r|}{ARIMA} & 18.9 & 19.5 & 20.3 & 20.8 & 21.0 & 20.6 & 19.2 & 18.2 & 18.0 & 17.8 & 17.7 & 17.7 & 18.0 & 18.4 & 18.6 & 18.8 & 19.0 & 18.9 & 19.1 & 19.2 \\
\multicolumn{1}{r|}{Informer} & 0.0 & 0.0 & 0.0 & 0.0 & 0.0 & 0.0 & 0.1 & 0.1 & 0.1 & 0.1 & 0.1 & 0.1 & 0.2 & 0.2 & 0.3 & 0.4 & 0.5 & 0.5 & 0.5 & 0.6 \\
\multicolumn{1}{r|}{DeepPrefetcher} & 74.0  & 76.2  & 79.8  & 81.4  & 82.1  & 82.1  & 81.4  & 80.8  & 80.9  & 80.9  & 80.9  & 81.4  & 82.8  & 83.9  & 84.9  & 85.6  & 86.2  & 86.5  & 86.8  & 87.0  \\
\multicolumn{1}{r|}{Delta-LSTM} & 75.9 & 77.9 & 81.1 & 82.5 & 83.0 & 82.9 & 82.1 & 81.5 & 81.4 & 81.4 & 81.4 & 81.9 & 83.1 & 84.3 & 85.2 & 85.9 & 86.5 & 86.8 & 87.1 & 87.3 \\ \hline
\multicolumn{1}{r|}{SGDP} & \textbf{80.0} & \textbf{82.5} & 85.9 & 87.7 & 88.5 & 88.8 & 88.5 & 88.3 & 88.4 & 88.5 & 88.5 & 89.0 & 89.7 & 90.1 & 90.3 & 90.4 & 90.5 & 90.6 & 90.7 & 90.8 \\
\multicolumn{1}{r|}{SGDP$_l$} & 79.0 & 81.5 & 84.9 & 86.7 & 87.5 & 87.7 & 87.4 & 87.2 & 87.3 & 87.4 & 87.4 & 87.8 & 88.3 & 88.6 & 88.7 & 88.9 & 89.0 & 89.1 & 89.3 & 89.3 \\
\multicolumn{1}{r|}{SGDP$_p$} & 73.6 & 76.1 & 80.2 & 82.4 & 83.7 & 84.1 & 83.7 & 83.5 & 83.7 & 83.9 & 84.0 & 84.6 & 85.3 & 85.7 & 86.1 & 86.4 & 86.8 & 86.9 & 87.2 & 87.3 \\ \hline \hline
\end{tabular}

        }
	\label{tab:src1_2}
\end{table*}

\end{document}